\makeatletter
\def\input@path{{template_aoas/}}
\makeatother

\documentclass[aoas]{imsart}

\RequirePackage{amsthm,amsmath,amsfonts,amssymb}
\RequirePackage[authoryear]{natbib}
\usepackage[colorlinks,citecolor=blue]{hyperref} 
\usepackage{xcolor}
\usepackage{enumitem}
\usepackage[ruled,vlined]{algorithm2e}
\usepackage{svg}

\usepackage{tikz}
\usetikzlibrary{calc, fit,backgrounds,matrix,positioning,arrows.meta,patterns}
\usetikzlibrary{calc,decorations.pathreplacing}

\startlocaldefs
\theoremstyle{plain}
\newtheorem{theorem}{Theorem}
\newtheorem{lemma}[theorem]{Lemma}
\theoremstyle{definition}
\newtheorem{assumption}{Assumption}
\usepackage{sty/my_smile}
\usepackage{sty/macro}

\endlocaldefs

\begin{document}

\begin{frontmatter}
\title{Matrix Factorization-Based Solar Spectral Irradiance Missing Data Imputation with Uncertainty Quantification}

\begin{aug}


\author[A]{\fnms{Yuxuan}~\snm{Ke}
    \ead[label=e1]{yuxuank@umich.edu}
    \orcid{0009-0003-1119-0807}
    },
\author[B]{\fnms{Xianglei}~\snm{Huang}
    \ead[label=e2]{xianglei@umich.edu}
    \orcid{0000-0002-7129-614X}
    },
\author[C]{\fnms{Odele}~\snm{Coddington}
    \ead[label=e3]{Odele.Coddington@lasp.colorado.edu}
    \orcid{0000-0002-4338-7028}
    }
\and
\author[A]{\fnms{Yang}~\snm{Chen}
    \ead[label=e4]{ychenang@umich.edu}
    \orcid{0000-0002-9516-8134}
    }

\address[A]{Department of Statistics,
University of Michigan \printead[presep={,\ }]{e1,e4}}

\address[B]{Department of Climate and Space Science, University of Michigan \printead[presep={,\ }]{e2}}

\address[C]{Laboratory For Atmospheric And Space Physics,      University of Colorado Boulder \printead[presep={,\ }]{e3}}

\end{aug}

\begin{abstract}
The solar spectral irradiance (SSI) depicts the spectral distribution of solar energy flux reaching the top of the Earth's atmosphere. 
Daily SSI measurements constitute a matrix with spectrally (rows) and temporally (columns) resolved solar energy flux measurements. 
The most recent SSI measurements have been made by NASA's Total and Spectral Solar Irradiance Sensor-1 (TSIS-1) Spectral Irradiance Monitor (SIM) since March 2018. 
This data has considerable missing data due to both random factors and instrument downtime, a periodic trend related to the Sun's cyclical magnetic activity, and varying degrees of correlation among the spectra, some approaching unity.

We propose a low-rank matrix factorization method for SSI reconstruction that incorporates autoregressive temporal regularization, periodic spline detrending, and cross-spectral covariance information. 
The method is implemented as a two-stage procedure designed to address scattered missingness and extended downtime missingness, respectively, and is fitted using efficient alternating optimization algorithms. 
We further accompany the reconstructed SSI values with a distribution-free interval estimation procedure based on conformal prediction. 
Through synthetic experiments and real-data analyses, we compare this method with Gaussian process regression, linear time series smoothing, and existing matrix-completion approaches in terms of imputation accuracy, interval coverage, interval length, and computational efficiency. 
The results show that exploiting the periodic, temporal, and cross-spectral structure of SSI substantially improves reconstruction performance and yields calibrated uncertainty intervals, producing a reconstructed SSI data product suitable for downstream climate science studies.
\end{abstract}


\end{frontmatter}

\section{Introduction} 
\label{sec:intro}
The solar spectral irradiance (SSI, $\text{W}\cdot \text{m}^{-2}\cdot \text{nm}^{-1}$) is the spectral distribution of solar energy flux reaching the top of the Earth's atmosphere. It is typically measured across a wavelength range from 200 to 2400 nm, spanning from ultraviolet to near-infrared regions of the electromagnetic spectrum. It exhibits periodicity driven by the 11-year solar magnetic cycle, the (roughly) 27-day solar rotation cycle, and the annual periodic variation in the Sun–Earth position \citep{coddington_solar_2016, lean_magnetic_1998}. It also exhibits varying degrees of correlation among the spectra, with some approaching unity. In SSI measurements, the wavelength range is divided into thousands of channels, each corresponding to a specific wavelength interval. 



As the sole external energy input to the climate system, the incoming solar energy flux is one of the most fundamental quantities to measure and monitor in climate science  \citep{ohring_satellite_2005, haigh_sun_2007}.
For example, the global surface temperature can be estimated using multiple linear regression with total solar irradiance as one of the predictors \citep{amdur_global_2021}.
In addition, the SSI is the upper boundary condition in terms of modeling the shortwave radiative transfer process in the atmosphere, a process indispensable for numerical weather forecasting, climate simulation, and air pollution modeling that involves photochemical reactions. As a result, high-quality daily SSI data are a prerequisite for historical climate simulations, such as those conducted for the Intergovernmental Panel on Climate Change \citep{funke_definition_2024, matthes_solar_2017}. 

\begin{figure}[t]
    \centering
    \begin{subfigure}[b]{\linewidth}
        \includegraphics[width=\textwidth]{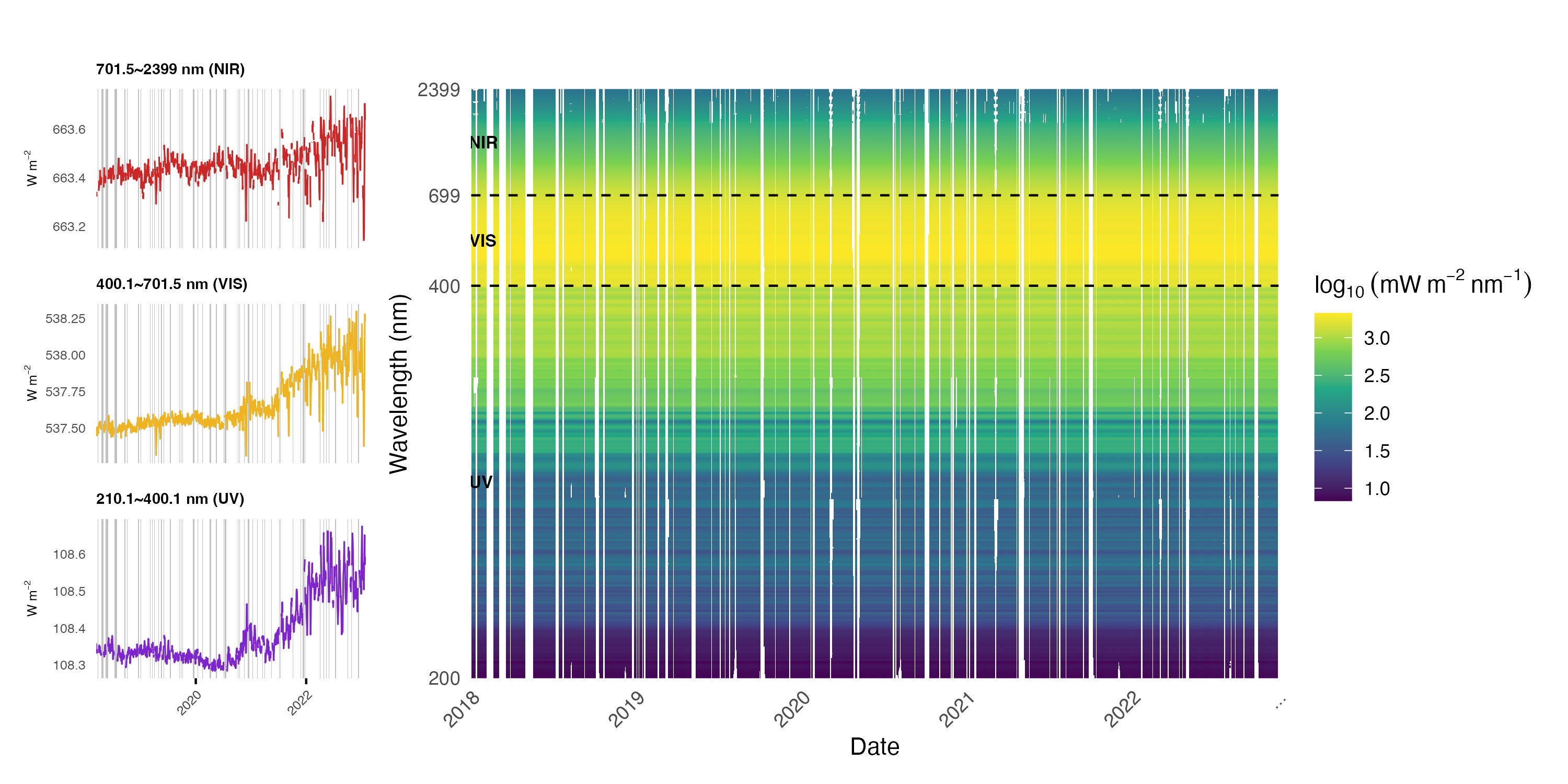}
    \end{subfigure}
    \caption{Solar spectral irradiance (SSI) data. \textbf{Left:} Time series of band-integrated irradiance over the full observation period for the near-infrared (NIR), visible (VIS), and ultraviolet (UV) spectral bands. Each panel integrates irradiance over its corresponding wavelength range. Gray vertical stripes indicate downtime missingness. \textbf{Right:} The full SSI matrix, with wavelength on the y-axis and date on the x-axis, colored by log-scaled irradiance.}
    \label{fig:ssi}
\end{figure}

The primary objective in keeping a continuous SSI record is to recover the missing entries in it (see Figure~\ref{fig:ssi}). We observed a moderate missing percentage of 10\% - 20\% in TSIS-1 SSI data. However, the missingness pattern is not uniformly random, as shown by the blank columns in Figure~\ref{fig:ssi}, which we call \textit{downtime} missingness when none of the wavelength channels is observed. This type of missingness usually occurs due to the instrument being temporarily blocked by the reinstallations of other experiments taking place on the International Space Station.
The left panel displays the band-integrated irradiances at four distinct wavelength bands, ranging from the ultraviolet to near-infrared regions, and spectrally integrate. 
In climate science, SSI reconstruction is commonly obtained by regression on solar activity proxies, for example, the f10.7 index, sunspot number, and Ca K intensity \citep{coddington_solar_2016, kakuwa_investigation_2022, kopp_impact_2016, yeo_empire_2017, ermolli_recent_2013}. 
Solar irradiance models that use only the direct observations of these solar magnetic features as the predictors of the regression analysis are called \textit{proxy models} or \textit{empirical models}. 
In contrast, \textit{semi-empirical models} incorporate both these direct observations and the intensity spectra from physical models of the Sun’s atmosphere.
Both kinds of methods need external data that correlates with solar irradiance to interpolate and extrapolate the missing SSI values. 
Prevalent distributionally agnostic statistical imputation methods, such as matrix completion \citep{fazel_matrix_2002, candes_exact_2009, hastie_matrix_2015}, multiple imputation by chained equations (MICE, \citealp{vanbuuren_flexible_1999}), and hot deck \citep{david_alternative_1986}, are powerful and require no external data, but they are ineffective under downtime missingness.
Note that SSI is mainly valued as a fundamental \emph{climate data record} (CDR) that supports downstream questions about solar variability rather than as single-time ``best guess''. The downstream signals of interest are typically expressed, for example, through band-integrated irradiance and trend estimates. Therefore, we need to control both bias and variance for the SSI reconstruction, since small errors at the spectral level can propagate into substantial ones in quantities such as band-integrated irradiance and long-term trends.
These requirements align with established evaluation criteria for climate data records, with an absolute accuracy of $0.2\%$, a relative precision of $0.01\%$ \citep{coddington_climate_2015, coddington_solar_2016}, and a $1$-sigma uncertainty goal of 0.2\% (maximum allowable threshold 1\%) \citep{richard_si-traceable_2020}. The reliability of associated uncertainty quantification is also emphasized. These considerations motivate our focus on methods that achieve strong imputation accuracy while providing robust uncertainty quantification that remains reliable under tructured missingness and an agnostic data generation distribution.

We propose a method that tackles the reconstruction problem from the matrix completion perspective, because of its flexibility, simplicity, and accuracy for matrices with large missing proportions \citep{candes_exact_2009}. 
We refine the matrix completion by introducing periodicity and temporal smoothness through penalization terms in the loss function. We also incorporate cross-spectral covariance to leverage information shared across contemporary spectral channels. This imputation method is data-driven and uses only the observed SSI, which is useful if other proxies of solar activity are unavailable. It keeps minimal assumptions on the data generation process while ensuring internal consistency specific to the measurement instrument.
We also provide the prediction intervals of the point estimates by modifying conformalized matrix completion \citep{gui_conformalized_2023}, which achieves a near-nominal level of empirical coverage, with $1$-sigma uncertainty (i.e., $1/1.96$ of relative half-lengths of the 95\% interval) less than or approximately 1\%, the maximum allowable threshold.
We compare the proposed algorithm against model-based imputation techniques such as Gaussian process kriging and linear time series smoothing. 
Across both synthetic experiments and real TSIS-1 SSI analyses, the proposed method consistently achieves strong imputation accuracy while maintaining reliable, near-nominal uncertainty quantification under the missingness patterns, yielding reconstructed SSI records that preserve scientifically relevant variability and uncertainty for downstream climate applications. The analyses also reveal how the unique characteristics of SSI contribute to imputation accuracy and emphasize the importance of leveraging this information. The  \href{https://deepblue.lib.umich.edu/data/concern/data_sets/5d86p133v}{synthetic benchmark data} and \href{https://deepblue.lib.umich.edu/data/concern/data_sets/rx913r011}{reconstructed SSI data product} are publicly available, together with open-source code at the \href{https://github.com/walnut-7/siap.git}{GitHub repository}, to support reproducibility and facilitate further scientific use. 


The paper is organized as follows. Section~\ref{sec:related works} reviews the literature on missing data imputation and existing methods. 
Section~\ref{sec:method} introduces our proposed SoftImpute with autoregressive regularization and periodic smoothing (SIAP) algorithm, together with its convergence rate and asymptotic properties. 
Section~\ref{sec:cp} describes the uncertainty quantification method.
Section~\ref{sec:simulation} and~\ref{sec:realdata} present and evaluate the results of the simulation study and the SSI data reconstruction, respectively. 
Section~\ref{sec:conclusion} discusses the findings and summarizes our contributions. 

\section{Relevant Literature}
\label{sec:related works}

\subsection{Statistical Analysis with Missing Data} 
\label{sec:stats analysis with missing data}

The validity of inferring the missing values from the observed ones relies on the fact that the missingness mechanism is not related to the data distribution, formally known as \textit{ignorability}. 
Let $X\in\RR^{m\times n}$ be the data matrix observed on $\Omega$, and $M\in\RR^{m\times n}$ the missingness indicator matrix, with $M_{ij}= 1$ if $(i,j) \in \Omega$ and $0$ otherwise.
The observed data are denoted by $\cO(X,M)$, which selects from $\vvec(X)$ the subvector consisting solely of those entries whose corresponding components in $\vvec(M)$ are equal to 0. The distributions of $X$ and $M$ are governed by data generation mechanism $p(X|\theta)$, and missingness mechanism $p(M|X;\psi)$, respectively. The parameters $(\theta,\psi)$ can be estimated by maximizing the \textit{joint likelihood} of the full observed data, $\rbr{\cO(X,M), M}$ \citep{little_chapter_2002}
\[
\ p(\cO(\tilde X,\tilde M), \tilde M|\theta,\psi) = \int p(X|\theta) p(\tilde M|X;\psi) \1\cbr{\cO(X,\tilde M)=\cO(\tilde X,\tilde M)} dX,
\]
where $\tilde X, \tilde M$ denote the observed realizations of the random variables $X$ and $M$.
In contrast, assuming \textit{ignorability} of the missingness mechanism \citep{azur_multiple_2011, bashir_handling_2018, josse_missmda_2016, schneider_analysis_2001} allows $\theta$ to be estimated using the \textit{likelihood ignoring the missing-data mechanism}
\begin{equation}
    \int p(X|\theta) \1\cbr{\cO(X,\tilde M)=\cO(\tilde X,\tilde M)} dX.
    \label{eq:likelihood ignoring the missing-data mechanism}
\end{equation}
For example, many model-based imputation methods, including the expectation maximization (EM) algorithm \citep{dempster_maximum_1977}, infer $\theta$ from $p(X|\theta)$.

Ignorability is appropriate when the data are missing at random (MAR, see \citealp[Definition 1]{rubin_inference_1976} and \citealp[Definition 1]{seaman_what_2013}) and the parameter $\theta$ is \textit{distinct} from $\psi$ \citep[Definition 3]{rubin_inference_1976}. Intuitively, MAR means that the probability of observing $M=\tilde M$ given the data $X$ does not depend on the value of its unobserved part.
Technically, ignorability implies that maximizing Eq.~\eqref{eq:likelihood ignoring the missing-data mechanism} is equivalent to maximizing the joint likelihood.

In the TSIS-1 SSI data, the occurrence of missing values is not related to the actual irradiance measurements themselves. Solar irradiance is generally stable over time, and the instruments are engineered to reliably capture the anticipated range of values. The scattered gaps in the data typically result from calibration issues or elevated levels of space-related noise.
Downtime missingness typically occurs when the instruments are temporarily blocked by other activities on the International Space Station. Both types of missingness are independent of $X$, supporting the MAR assumption.
Therefore, the missing mechanism can be considered as ignorable. This allows us to infer the model parameter $\theta$ from $p(X|\theta)$. In the next section, we review some existing approaches that build on this assumption.

\subsection{Existing Approaches}



There are three major methodological approaches in the literature that could be suitable for the problem of missing-data inference and uncertainty quantification for SSI: loss-function-based methods from matrix completion, likelihood-based time series models, and nonparametric smoothing methods. We provide further details on each category in this section.

Matrix completion is the task of recovering the missing entries in a large data matrix. It was first made famous by Netflix's 2006 open competition to improve its movie recommendation system. Besides recommendation systems \citep{rennie_fast_2005, mongia_matrix_2019}, matrix completion also has wide applications in image restoration \citep{li_regularised_2022}, data integration \citep{cai_structured_2016}, and longitudinal data analysis \citep{kidzinski_modeling_2024}. 
The matrix low-rank completion technique is based on the intuition that the degree of freedom of the essential signal of the matrix is much smaller than the total number of entries, thus assuming the underlying signal matrix to be low-rank. 
Let $X$ denote the observed matrix and $\Omega=\{(i,j):x_{ij}\text{ is observed}\}$ the observed entries index set. 
There are two main streams in the matrix completion research: 1) those that assume the observed matrix is noise-free, and minimize $\rank(Y)$ with respect to $Y$ subject to $\cP_\Omega(X)=\cP_\Omega(Y)$;
and 2) those that assume the observed matrix is a noisy version of the underlying low-rank signal matrix, and minimize the rank subject to $h\rbr{\cP_\Omega(Y-X)} \leq \delta$, where $h(\cdot)$ is a specific error function and $\delta>0$ is a regularizing parameter.
One common choice of $h$ is the sum of squared error \citep{mazumder_spectral_2010}. In robust matrix completion, $h$ is specified as a robust loss function such as Huber loss \citep{wong_matrix_2017}.

Several ideas have been proposed to relax the rank restriction, including nuclear norm minimization (NNM \citealp{candes_exact_2009, fazel_matrix_2002, candes_power_2010, recht_guaranteed_2010}), and low Frobenius norm factorization \citep{rennie_fast_2005, hastie_matrix_2015}. 
The NNM problem can be solved by 1) semi-definite program (SDP) solvers \citep{srebro_maximum-margin_2004} which become expensive when the matrix is large, or 2) \textit{singular value thresholding} (SVT) which iteratively shrinks the singular values of the matrix \citep{cai_singular_2010}, or 3) by adopting an EM idea \citep{mazumder_spectral_2010}. 
The low Frobenius norm factorization problem, 
\begin{equation}
    \begin{aligned}
        \min_{A\in\RR^{m\times r},B\in\RR^{n\times r}} &\left\{ \frac{1}{2}\fnorm{\cP_\Omega(X-AB^{\top})}^2 + \frac{\lambda}{2}(\fnorm{A}^2 + \fnorm{B}^2)\right\},
    \end{aligned}
    \label{eq:matrix factorization}
\end{equation}
due to its smoothness and bi-convexity, can be solved by gradient-based methods \citep{rennie_fast_2005} or \textit{alternating minimization} \citep{jain_low-rank_2013, hastie_matrix_2015}. 
Specifically, \citet{hastie_matrix_2015} propose the \texttt{SoftImpute-ALS} algorithm, adopting both the alternating minimization idea and the EM-type idea, thus making it one of the most time-efficient algorithms. It iteratively fills in the missing entries with the current estimate $\hat X$ and solves the low-Frobenius-norm factorization.
Problem~\eqref{eq:matrix factorization} has also been examined from a Bayesian perspective \citep{salakhutdinov_bayesian_2008, nakajima_theoretical_2011}, as it is equivalent to maximizing the posterior of $A,B$ under Gaussian priors.

Recent matrix completion methods modify the conventional methods to accommodate settings with non-uniform missingness and temporal dependence. \citet{xiong_temporal_2010, rallapalli_exploiting_2010} apply a graph-based regularizer to model the dependence. \citet{yu_temporal_2016, chen_lowrank_2022} and \citet{chen_back_2024} incorporate temporal dependence by imposing an autoregressive (AR) regularizer either on the latent temporal factors or on the observed data space. Some also account for seasonal behavior by folding the matrix into a tensor \citep{chen_lowrank_2022} or by incorporating periodic autoregressive components \citep{chen_back_2024}.


Estimating the missing values in multivariate time series can also be tackled from a vector time series perspective. 
\cite{davis_sparse_2016, wilms_interpretable_2017, nicholson_high_2020, wilms_sparse_2023} provide effective methods to model high-dimensional linear time series, but are unable to deal with the missing data.
\cite{liu_ivar_2014} uses one-step-ahead predictions by the vector autoregressive (VAR) model to impute missing data.
\cite{holmes_marss_2012, bashir_handling_2018} further combines the VAR and expectation maximization (EM, \citealp{dempster_maximum_1977}) idea and alternates between estimating the missing values and the model parameters.
An extension to VAR that considers a more general cross-sectional correlation and low-dimensional structure is the multivariate autoregressive state-space model (MARSS, \citealp{holmes_marss_2012, kohn_fixed_1983, aoki_chapter_2013b, anderson_chapter_1979}). 

An alternative perspective treats $X$ as a 2-dimensional surface, enabling Gaussian process (GP, \citealp{murphy_chapter_2022, williams_prediction_1998}) kriging to model dependencies among entries. When $|\Omega|=O(mn)$, the primary computational cost arises from large matrix inversion, which scales as $O(m^3n^3)$ in a naive implementation. Refinements were done by reducing the intrinsic dimensionality of the covariance structure. Some consider reducing the complexity in the covariance matrix by, for example, the knot-based methods that assume the GP likelihood is dominant by $r \ll mn$ latent knots \citep{snelson_sparse_2005, gelfand_spatial_2012}, or directly introducing sparsity in the covariance matrix by assuming independence across geographical regions \citep{sang_full_2012}, or by establishing low-rank kernel functions \citep{nychka_multiresolution_2015, wesel_tensorbased_2023, chen_linearcost_2023}. Others, such as Vecchia's approximation \citep{vecchia_estimation_1988}, nearest neighbor GP (NNGP, \citealp{datta_hierarchical_2016, guinness_permutation_2018, guinness_gaussian_2021}), and meshed GP (MGP, \citealp{peruzzi_highly_2022}), assume sparsity in the precision matrix, motivated by the directed acyclic graphical (DAG) representation of the GP likelihood. Furthermore, \citet{gardner_gpytorch_2018} improve the computational efficiency using the conjugate gradient methods. 





\section{SIAP: Utilizing Spectral and Temporal Information for Imputation}
\label{sec:method}

\begin{figure}[t]
\centering
\begin{tikzpicture}
\node[draw=cyan!70!black, thick, rounded corners, fill=cyan!10, fit={(3.1, -2) (3.7, -1.53)}, inner sep=2pt] (sig) {};

\node[draw=blue!70!black, thick, rounded corners, fill=blue!10, fit={(6.6, -2) ($(6.6, -2)+(0.6, 0.47)$)}, inner sep=2pt] (mu1) {};

\node[draw=blue!70!black, thick, rounded corners, fill=blue!10, fit={(9.3, -4.7) ($(9.3, -4.7)+(0.6, 0.47)$)}, inner sep=2pt] (mu2_1) {};
\node[draw=blue!70!black, thick, rounded corners, fill=blue!10, fit={(7.8, -6.9) ($(7.8, -6.9)+(0.4, 0.47)$)}, inner sep=2pt] (mu2_2) {};
\node[draw=blue!70!black, thick, rounded corners, fill=blue!10, fit={(10, -6.9) ($(10, -6.9)+(0.4, 0.47)$)}, inner sep=2pt] (mu2_3) {};

\node at ($(sig.north) + (0.1,0.15)$) [anchor=south, text=cyan!40] {\small principal submatrix of $\Sigma$};
\node at ($(mu1.north) + (0.1,0.15)$) [anchor=south, text=blue!80!black] {\small cubic spline function};
\node (mu2_lab) at (11, -5.5) [anchor=west, text=blue!80!black] {\small cubic spline function};

\draw[->, thick, blue] (mu2_1.south) -- ($(mu2_lab.west) + (0,0.1)$);
\draw[->, thick, blue] (mu2_2.north) -- (mu2_lab.west);
\draw[->, thick, blue] (mu2_3.north) -- ($(mu2_lab.west) + (0,-0.1)$);

\node[anchor=north west] (eq1) at (0,0) {
\begin{minipage}{0.92\textwidth}
\textbf{Step 1: Handling periodicity and cross-spectra correlation (Section~\ref{sec:step 1})}
\vspace{.67cm}
\[
\min_{A, B, \Theta, \Sigma} 
\textcolor{red}{\underbrace{\color{black}\sum_{t} 
\left\| 
\Sigma_t^{-1/2}
\left( \bx_{\text{obs}, t} - A_{\text{obs}} \bb_t - {\bmu_{\text{obs}}(t)} \right) 
\right\|^2
}_{
\begin{minipage}{0.6\textwidth}
    \centering
    \small
    \linespread{0.9}\selectfont
    generalized squared error on observed entries
\end{minipage}
}} 
+ \textcolor{gray}{\underbrace{\color{black}\lambda \left( \|A\|_F^2 + \|B\|_F^2 \right)}_{
\begin{minipage}{0.3\textwidth}
    \centering
    \small
    \linespread{0.9}\selectfont
    $\ell_2$ regularization
\end{minipage}
}}
\]
\end{minipage}
};

\node[anchor=north west] (eq2) at (0,-3.5) {
\begin{minipage}{0.92\textwidth}
\textbf{Step 2: Latent space autoregressive regularization (Section~\ref{sec:step 2})}
{
\[
\begin{aligned} 
\min_{A, B, \Theta} &
\textcolor{red}{\underbrace{\color{black}\left\| P_{\Omega_2} (X^{(1)} - A B^T) \right\|_F^2}_{
\begin{minipage}{0.2\textwidth}
    \centering
    \small
    \linespread{0.9}\selectfont
    \textcolor{red}{squared error on observed entries}
\end{minipage}
}}
+ \textcolor{gray}{\underbrace{\color{black} \lambda_1 \|A\|_F^2 
+ \lambda_2 \left\| B_{1:p} - {\mu_{1:p}} \right\|_F^2}_{
\begin{minipage}{0.3\textwidth}
    \centering
    \small
    \linespread{0.9}\selectfont
    $\ell_2$ regularization
\end{minipage}
}} \\
& + 
\textcolor{yellow!80!black}{\underbrace{\color{black} \alpha \sum_{j=1}^{n-p} \left\| 
{\bb_{j+p}} - \sum_{k=1}^{p} \Gamma_k \left( {\bb_{j+p-k}} - {\bmu(j + p - k)} \right)
- {\bmu(j + p)} 
\right\|^2}_{
\begin{minipage}{0.3\textwidth}
    \centering
    \small
    \linespread{0.9}\selectfont
    AR regularization
\end{minipage}
}}
\end{aligned}
\]
}
\end{minipage}
};

\end{tikzpicture}
\caption{Overview of the algorithms' loss functions.} 
\label{fig:loss function diagram}
\end{figure}

\begin{figure}[tb]
    \centering
    \begin{tikzpicture}[
      font=\small,
      row sep=0.5pt,
      process/.style={rectangle, draw, rounded corners, align=left, 
      minimum height=2em, inner sep=5pt, text width=4cm},
      circlebox/.style={circle, draw, minimum size=8mm, align=center},
      arrow/.style={-{Latex[length=2mm]}, thick},
      node distance=0.3cm and 1cm
    ]
    
    \node[circlebox] (step1circle) {I};
    \node[align=left, right=0.2cm of step1circle, anchor=west] (step1label) {\textbf{SIAP Step 1}};
    \node[process, below=of step1label] (step1_1) {
    \setlength{\baselineskip}{15pt} 
    (1) Detrending: periodic cubic spline in the observed space.\\ 
    \begin{minipage}{\textwidth}
        \centering
         {$\Theta^T \Phi^T$}
    \end{minipage}
    };

    \node[coordinate, below=1.7cm of step1_1] (dummycenter) {};
    \node[process, left=0.15cm of dummycenter, text width=2cm] (step1_2) {
    \setlength{\baselineskip}{15pt}
    (2) Low-rank matrix factorization.\\ 
    \begin{minipage}{\textwidth}
        \centering
         {$AB^T$}
    \end{minipage}
    };
    \node[process, right=0.15cm of dummycenter, text width=2cm] (step1_3) {
    \setlength{\baselineskip}{15pt}
    (3) Cross-sectional covariance\\ 
    \begin{minipage}{\textwidth}
        \centering
         {$\Sigma$}
    \end{minipage}
    };

    \node[draw, thick, rounded corners, fit=(step1label) (step1_1) (step1_2) (step1_3), 
      inner sep=6pt] (step1group) {};

    \node[circlebox, right=7.9cm of step1circle] (step2circle) {II};
    \node[align=left, right=0.2cm of step2circle, anchor=west] (step2label) {\textbf{SIAP Step 2}};
    \node[process, below=of step2label] (step2_4) {
    \setlength{\baselineskip}{15pt}
    (4) Detrending: periodic cubic spline in the latent space.\\ 
    \begin{minipage}{\textwidth}
        \centering
         {$\| B - \Phi \Theta \|_F^2$}
    \end{minipage}
    };
    \node[process, below=0.3cm of step2_4] (step2_5) {
    \setlength{\baselineskip}{15pt}
    (5) Temporal smoothing: AR regularization in the latent space.\\ 
    \begin{minipage}{\textwidth}
        \centering
         {$\left\| \tilde{\bb}_t - \sum_{k=1}^p \Gamma_k \tilde{\bb}_{t-k} \right\|_F^2$}
    \end{minipage}
    };
    
    \node[draw, thick, rounded corners, fit=(step2label) (step2_4) (step2_5), 
      inner sep=6pt] (step2group) {};

    \node[process, right=0.6cm of step1group, text width=2cm] (bridge) {
    \setlength{\baselineskip}{15pt}
    Impute the scattered missingness with BLUE. Keep downtime missingness.\\
    };
    \draw[arrow] (step1group.east) -- (bridge.west) node[midway, above] {\footnotesize };
    \draw[arrow] (bridge.east) -- (step2group.west) node[midway, above, sloped] {\footnotesize };
    \end{tikzpicture}
    \caption{Illustration of \texttt{SIAP}. Step~1 jointly estimates the observed space detrending (module 1), low-rank matrix factorization (module 2), and cross-sectional covariance (module 3). Step~2 models the temporal dynamics where module 4 is latent space detrending and module 5 is temporal smoothing using AR regularization. $\tilde B = (\tilde\bb_1,\ldots,\tilde\bb_n)^{\top} = B - \Phi\Theta$.}
    \label{fig:siap modules}
\end{figure}

To address the periodic trend, spectral dependency, and downtime missingness of SSI data, we propose a two-step algorithm called the Soft-Impute Algorithm with Autoregressive regularization and Periodic spline detrending (SIAP), as illustrated in Figure~\ref{fig:loss function diagram} and Figure~\ref{fig:siap modules}. 
We (1) employ periodic splines to standardize the data to effectively model the global trend, (2) implement AR regularization to exploit local temporal information, and (3) implement a cross-sectional variance-covariance structure to leverage spectral information. 
%
We address the scattered missingness and downtime missingness separately in each step. 
In the first step, we focus on using the cross-sectional variance-covariance structure and splines detrending in the \textit{observed space} to impute the scattered missingness. This entails applying an EM algorithm to estimate the missing values as well as the model parameters. We estimate the downtime missingness in SSI in the second step by applying AR regularization and spline fitting to the $B$ matrix, given the estimated values for the scattered missingness from the first step. 


\subsection{Step 1: Handling Periodicity and Cross-spectra Correlation} 
\label{sec:step 1}

Let $\bx_t$ and $\bb_t$ denote the $t$-th column of $X$ and $B^\top$, respectively.
For SSI data, because $\EE\bx_t$ varies with time $t$, we model it by function $\bmu(t)$, which implies
\[
\bx_t|A,B\overset{\mathrm{i.i.d}}\sim \cN(A\bb_t+\bmu(t),\Sigma),\ t=1,\ldots,n,
\]
where $\Sigma$ suggests a general covariance structure. 
As a comparison, since the first term in Eq.~\eqref{eq:matrix factorization} vanishes when $\bx_t$ is entirely missing, $\bb_t$ will shrink to 0. Therefore, the original matrix completion method is extremely sensitive to data preprocessing.

\begin{figure}[t]
\centering
\begin{tikzpicture}[
    box/.style={draw, thick},
    bigbox/.style={draw, thick, minimum width=4cm, minimum height=3cm},
    smallcol/.style={draw, thick, minimum width=1cm, minimum height=3cm}
]

\node[bigbox] (X) {$X$};

\node[draw, thick, minimum width=.5cm, minimum height=3cm,
      fill=red!30] (xt) at ([xshift=1cm]X.center) {$x_t$};


\draw[->, thick] ([yshift=.3cm]X.north west) -- ([yshift=.3cm]X.north east);
\node at ([yshift=.5cm]X.north) {Time};

\node[rotate=90] at ([xshift=-.5cm]X.west) {wavelength};

\node[right=.5cm of X] (approx) {$\approx$};

\node[draw, thick, minimum width=1.5cm, minimum height=3cm, right=.5cm of approx] (A) {$A$};

\node[draw, thick, minimum width=1.5cm, minimum height=0.5cm,
      fill=blue!30] at ([yshift=-.8cm]A.center) {$\ba_i^\top$};

\node[draw, thick, minimum width=4cm, minimum height=1.5cm, right=.5cm of A] (B) {$B$};

\node[draw, thick, minimum width=.5cm, minimum height=1.5cm,
      fill=red!30] at ([xshift=1cm]B.center) {$\bb_t$};


\node[above=0.5cm of B] {Time-dependent variables};

\draw[decorate,decoration={brace,amplitude=6pt}]
    ([yshift=0.2cm]B.north west) --
    ([yshift=0.2cm]B.north east);

\end{tikzpicture}
\caption{Matrix factorization representation of matrix $X$.}
\label{fig:matrix}
\end{figure}

We model $\bmu(t)$ using \textit{periodic cubic splines}, which leverage the periodicity of SSI and preserve more flexibility than sinusoidal curves.
Splines 
are smooth functions obtained from concatenating pieces of polynomials at a sequence of breakpoints called \textit{knots} \citep{wang_shaperestricted_2021, hastie_chapter_2009, gu_chapter_2013a}. 
The periodic basis splines are constructed from \textit{basis splines} (B-splines, \citealp{deboor_calculating_1972}) given the knots and the endpoints of a period \citep{wang_shaperestricted_2021}. 
With the basis splines $\{\phi_i\}_{i=1}^\kappa$ and $\bphi = (\phi_1,\ldots,\phi_\kappa)^{\top}$, $\bmu$ is parameterized by $\Theta^\top \bphi(t)$ where $\Theta = (\bm \theta_1, \bm \theta_2, \ldots ,\bm \theta_m)\in\mathbb R^{\kappa\times m}$ are the coefficients. When accounting for multiple periodicities, since the Sun exhibits several irradiance cycles, including the 27-day solar rotational cycle, 11-year solar activity cycle, and the year-round cycle caused by Earth’s orbit around the Sun, the corresponding bases are concatenated to form the basis matrix $\Phi$, which results in an additive model.

Our proposed algorithm optimizes the regularized negative marginal log-likelihood,
\begin{equation}
    -\sum_{j=1}^n\rbr{\log\left|\Sigma_{j,\cO_j\cO_j}^{-1}\right| - \left\|\Sigma_{j,\cO_j\cO_j}^{-\frac12}\rbr{\bx_{j,\cO_j}-\Theta^{\top} \bphi_{j,\cO_j}-A\bb_{j,\cO_j}}\right\|_2^2} + \lambda(\|A\|_F^2+\|B\|_F^2),
    \label{eq:1 loss}
\end{equation}
where subscript $\cO_j$ represents the observed entries in $\bx_j$. Similarly, we let $\cM_j$ denote the missing entries. For any vector $\bx$ and its observed entries $\cM$ and missing entries $\cO$, define its \textit{missing part} as
$\bx_{\cM} = (x_{k_1},\ldots,x_{k_{l}})\ ,{k_1}<\ldots<k_{l} \in \cM$, which extracts the missing subvector of $\bx$. Similarly, the \textit{observed part} $\bx_{\cO}$ is the observed subvector.
Accordingly, the submatrix $\Sigma_{j,\cM_j\cO_j}$ consists of the rows corresponding to the missing entries of $\bx_j$ and columns corresponding to the observed entries of $\bx_j$. $\Sigma_{j,\cO_j\cO_j}$ and $\Sigma_{j,\cM_j\cM_j}$ are the observed and missing principal submatrices of $\Sigma$, respectively.


The model parameters are estimated through the regularized expectation conditional maximization (ECM) \citep{dempster_maximum_1977, meng_maximum_1993, yi_regularized_2015} algorithm. The parameters $A,\ B,\ \Theta$ and $\Sigma$ are updated in an alternating fashion, each of which has a closed-form solution. 
We consider a generalized \textit{spiked covariance structure} \citep{johnstone_distribution_2001} $\Sigma = \Lambda + LL^{\top}$,
where $\Lambda$ is a diagonal matrix and $L$ is a tall matrix. This ensures a positive definite $\Sigma$ estimate even when $n<m$. It also provides an efficient estimate of the precision matrix by using the Woodbury matrix identity.
The updates for $\Lambda$ and $L$ are jointly obtained using EM under the latent variable model $\tilde \bx_j=\bx_j-(A\bb_j+\Theta^{\top}\bphi_j)$, $\bz_j\sim \cN(0,I)$, $\tilde \bx_j|\bz_j\sim \cN(L\bz_j,\Lambda)$, with $\bz_j$ being the hypothetical latent variable. 


Algorithm~\ref{alg:1} summarizes the steps of the proposed method, where $\conde{\cdot}$ denotes the conditional expectation given $X_{\cO},A,B,\Sigma,\Theta$.
The E-step involves computing all the first- and second-order condition expectations, which are listed in the supplement.
In practice, we initialize $\Lambda_1=10^{-4}\cdot I_m$, $\vvec(L_1) \sim \cN(\bm 0,10^{-4}\cdot I_{mr_L})$, where $r_L << m$ is the number of columns in $L_1$. 
$\Theta_1$ is initialized through spline fitting on $X$.
Then $A_1,\ B_1$ are obtained by applying the \texttt{SoftImpute-ALS} algorithm to $X-\Theta_1^\top\Phi^\top$.

\begin{algorithm}[t]
\SetAlgoLined
\caption{\texttt{SIAP}: Step 1}
\label{alg:1}
\textbf{Inputs: }Data matrix $X$, initialization $(A_1,B_1,\Theta_1,L_1,\Lambda_1)$, hyperparameters $r,\lambda,\rho$, tolerance $\epsilon,\epsilon'$.

\textbf{Outputs: }$(A^*,B^*,\Theta^*,\Sigma^*)$ which is an estimate of the minimizer of Problem~\eqref{eq:1 loss}, and $\conde{X}=\EE[X|\cP_\Omega(X),A^*,B^*,\Theta^*,\Sigma^*]$.


\begin{enumerate}
    \item For $k=1,\cdots,K,$ 
    \begin{enumerate}[topsep=0pt,itemsep=-1ex]
    \item Update conditional expectations, $\conde{X}$, $\conde{\bz_j \bz_j^{\top}}$, $\conde{\tilde\bx_j\tilde \bx_j^{\top}}$ and $\conde{\bz_j\tilde \bx_j^{\top}}$. 
    \item $B_{k+1} \leftarrow  (\conde{X^{\top}} - \Phi\Theta_k) \Sigma_k^{-1} A_k \rbr{A_k^{\top}\Sigma_k^{-1}A_k + \lambda I_r}^{-1}$,
    where $\conde{X^{\top}}=\conde{X}^{\top}$.
    \item $L_{k+1} \leftarrow (\sum_{j=1}^n \conde{\tilde \bx_j\bz_j^{\top}})(\sum_{j=1}^n  \conde{\bz_j\bz_j^{\top}})^{-1}$, and $\Lambda_{k+1} \leftarrow \diag\{\frac1n\sum_{j=1}^n[\conde{\tilde \bx_j\tilde \bx_j^{\top}}] - 2 L_{k+1} \conde{\bz_j\tilde \bx_j^{\top}} 
    + L_{k+1} \conde{\bz_j\bz_j^{\top}}L_{k+1}^{\top}\}$.
    \item $A_{k+1} \leftarrow (\conde{X} - \Theta_k^{\top}\Phi^{\top})B_{k+1}(B_{k+1}^{\top} B_{k+1}+\lambda I_r)^{-1}$.
    \item $\Theta_{k+1} \leftarrow (\Phi^{\top}\Phi)^{-1}\Phi^{\top}(\conde{X} - A_{k+1}B_{k+1}^{\top})^{\top}$.
    \item Break when $k=K$, or $\frac{\|A_{k+1}{B_{k+1}}^{\top} - A_k{B_k}^{\top}\|_F^2}{\|A_k{B_k}^{\top}\|_F^2}<\epsilon$ and $\frac{\fnorm{\Lambda_{k+1}-\Lambda_{k}}^2}{\fnorm{\Lambda_{k}}^2}<\epsilon'$.
\end{enumerate}
    \item $(A^*,B^*,\Theta^*,\Sigma^*)\leftarrow(A_{k+1},B_{k+1},\Theta_{k+1},\Sigma_{k+1})$.
\end{enumerate}
\end{algorithm}

\subsection{Step 2: Latent Space Autoregressive Regularization}
\label{sec:step 2}
Under the noisy assumption that $X=X^*+E$, many matrix completion methods aim to solve $X^*$. 
In contrast, our method directly targets $X$. After Step~1, the scattered missingness in $X$ is imputed by $\conde{X}$. 
Let $X^{(1)}$ denote the resulting partially imputed matrix, and $\Omega_1$ denote the index set of observed entries in $X^{(1)}$. 
The Step~2 algorithm solves the problem below:
\begin{equation}
    \begin{aligned}
        \min_{A,B,\Theta}& \underbrace{\|\cP_{\Omega_1}(X^{(1)}-AB^{\top})\|_F^2}_{\cL_1}
        + \underbrace{\lambda_1\|A\|_F^2 + \lambda_2\|B_{1:p}-\eta_{1:p}\|_F^2}_{\cL_2} \\
        & + \underbrace{\alpha\sum_{j=1}^{n-p}\|(\bb_{j+p}-\bmeta(j+p))-\sum_{l=1}^p\Gamma_l (\bb_{j+p-l} - \bmeta(j+p-l))\|^2}_{\cL_3},
    \end{aligned}
    \label{eq:2 loss}
\end{equation}
where $B_{1:p}=(\bb_1,\ldots,\bb_p)^\top$, and $\eta_{1:p}=(\bmeta(1),\ldots,\bmeta(p))^\top$.
At this stage, we let the matrix $B$ capture all the temporal dynamics, including the periodic trends, and model the trend of the latent process $\{\bb_t\}_{t>0}$ by $\bmeta(t)$.
In addition, we impose autoregressive (AR) regularization on $B$ instead of Frobenius norm regularization, represented by $\cL_3$ in Eq.~\eqref{eq:2 loss}, where $\cbr{\Gamma_l}_{l=1}^p$ are assumed diagonal. We refer to $p$ in Eq.~\eqref{eq:2 loss} as the maximum time lag of the AR regularization. The mean function $\bmeta$, similar to $\bmu$ in Step~1, is modeled using a periodic spline and parameterized as $\bmeta(t) = \Theta^{\top} \bphi(t)$.

Algorithm~\ref{alg:2} gives the outline. Similar to Step~1, the parameters are updated alternately. When $r>15$, we update $\bb_t$ sequentially for $t=1,\ldots,n$, solving a series of linear systems of dimensions $r$. This significantly increases the algorithm's scalability. Detailed derivations of the update equations are provided in the supplement.
The initialization points $A_1,\ B_1$ are obtained by applying \texttt{SoftImpute-ALS} to $X^{(1)}$. And $\Theta_1$ is initialized through spline fitting on $B_1$. Finally, we impute the downtime missingness by $A^{*}{B^*}^{\top}$.
For computational efficiency, $\cbr{\Gamma_l}$ are considered as hyperparameters instead of the estimated parameters. They are determined by the least-squares solution using the non-downtime columns of $(B_1-\Theta_1)^\top$.
Our sensitivity analysis shows that iteratively updating $\cbr{\Gamma_l}$ does not significantly increase imputation accuracy and may introduce unreliable artifacts.  

\begin{algorithm}[t]
\SetAlgoLined
\caption{\texttt{SIAP}: Step 2}
\label{alg:2}
\textbf{Inputs: }Data matrix $X^{(1)}$, initialization point $(A_1,B_1,\Theta_1)$, hyperparameters $r,p,\lambda_1,\lambda_2,\alpha$, tolerance $\epsilon$.

\textbf{Outputs: }$(A^*,B^*,\Theta^*)$ which is an estimate of the minimizer of Problem~\eqref{eq:2 loss}, and $\hat X$. 

\begin{enumerate}
    \item For $k=1,2,\cdots,K,$ 
    \begin{enumerate}[topsep=0pt,itemsep=-1ex]
    \item $\Theta_{k+1} \leftarrow (\Phi^{\top}\Phi)^{-1}\Phi^{\top}{X_\Theta^*}^{\top}A_k (A_k^{\top}A_k)^{-1}$, where $$X^*_\Theta=\cP_{\Omega_1} = \cP_{\Omega_1}(X^{(1)}-A_k B_k^{\top}) + A_k\Theta_k^{\top}\Phi^{\top}.$$
    \item Update $B_{k+1}$. 
    \item $A_{k+1} \leftarrow (B_{k+1}^{\top}B_{k+1} + \lambda_1 I_r)B_{k+1}^{\top}{X^*_A}^{\top}$, where $X^*_A = \cP_{\Omega_1}(X^{(1)})+\cP_{\Omega_1}^\perp(A_kB_{k+1}^{\top})$.
    \item Break the iteration when $k=K$ or $\frac{\|A_{k+1}{B_{k+1}}^{\top} - A_k{B_k}^{\top}\|_F^2}{\|A_k{B_k}^{\top}\|_F^2}<\epsilon$.
\end{enumerate}
    \item $(A^*,B^*,\Theta^*)\leftarrow(A_{k+1},B_{k+1},\Theta_{k+1})$. The final estimate of $X$ is $$\hat X \leftarrow \cP_{\Omega_1}(X^{(1)})+\cP_{\Omega_1}^\perp( A^* {B^*}^{\top}).$$
\end{enumerate}
\end{algorithm}

\subsection{Algorithm Convergence Analysis}
\label{sec:conv}
In this section, we show that the iterates of Algorithm~\ref{alg:2} converges to a stationary point of the loss function Eq.~\eqref{eq:2 loss}. 
We focus on Step~2, as the results for Step~1 follow similar principles. Proofs are provided in the supplement.



\begin{theorem}[Non-increasing loss function in Step 2] 
    \label{thm:2 decrease}
    Let $F_2$ denote the loss function in Problem~\eqref{eq:2 loss}. Then
    $F_2(A_k,B_k,\Theta_k) 
    \geq F_2(A_{k+1},B_{k+1},\Theta_{k+1}),\ k=1,2,\ldots.$
\end{theorem}

This theorem indicates that with each iteration of the algorithm, the value of the loss function decreases. Let $\Delta_k = F_2(A_k,B_k,\Theta_k) - F_2(A_{k+1},B_{k+1},\Theta_{k+1})$ denote the successive difference of the loss function values after the $k$-th iteration. Lemma~\ref{lem:2 Delta=0} connects this difference with the convergence of the iterates $\cbr{(A_k,B_k,\Theta_k)}_{k>1}$, enabling $\Delta_k$ to quantify the distance to the \textit{fixed point}. Let $\{\theta_j\}_{j\geq1}$ denote the sequence of iterations generated by the update $\theta_{k+1}\leftarrow f(\theta_k)$. Then $\theta_+$ is said to be a \emph{fixed point} if $f(\theta_+)=\theta_+$. 
Theorem~\ref{thm:2 up} establishes a convergence rate of $O(1/K)$, which means with $K=O(1/\epsilon)$ iterations, the algorithm would reach a point $(A_k,B_k,\Theta_k)$ such that $\Delta_k\leq \epsilon$.
\begin{lemma}
    \label{lem:2 Delta=0}
    $\Delta_k=0$ if and only if $(A_k,B_k,\Theta_k)$ is a fixed point of Algorithm~\ref{alg:2}.
\end{lemma}

\begin{theorem}[Upper Bound of $\Delta_k$]
    \label{thm:2 up}
    The decreasing sequence $\{F_2(A_k,B_k,\Theta_k)\}_{k\geq1}$ converges to $F_2^\infty \geq 0$ and $\lim_{k\rightarrow\infty}\Delta_k=0$. Furthermore,
    \begin{equation}
        \min_{1\leq k\leq K} \Delta_k\leq\frac{F_2(A_1,B_1,\Theta_1)-F_2^\infty}{K}.
    \end{equation}
\end{theorem}

We demonstrate in the supplement that the \textit{stationary point} (defined as a point with zero gradient of the loss function) of the loss function in \eqref{eq:2 loss} is equivalent to the fixed point of Algorithm~\ref{alg:2}. This leads to Theorem~\ref{thm:2 stationary point}, which summarizes the algorithmic convergence. 


\begin{theorem}[Convergence to stationary point]
    \label{thm:2 stationary point}
    For $\lambda_1,\lambda_2,\alpha>0$, every limit point of $\{A_k,B_k,\Theta_k\}_{k\geq 1}$ is a fixed point of the Algorithm, thus a first-order stationary point of Problem~\eqref{eq:2 loss}. 
\end{theorem}

\section{Uncertainty Quantification}
\label{sec:cp}

Besides point estimates, prediction intervals are also of great interest, both statistically and in the context of SSI reconstruction.  
Application-wise, they are needed because SSI completion is evaluated not only by how accurately missing values are filled in, but also by whether the reconstructed data support reliable downstream estimates, such as band-integrated irradiance and long-term trends. This aligns with SSI CDR requirements, which emphasize accuracy, precision, and explicit uncertainty targets \citep{coddington_climate_2015, coddington_solar_2016, richard_si-traceable_2020}.
Statistically, uncertainty typically comes from the data generation mechanism, and the missingness mechanism.
Accordingly, there are two approaches to formulating assumptions for quantifying uncertainty. The first is placing assumptions on the data generation mechanism. Both GP and MARSS models fall under this category. Assuming $X\sim\cX$ and a confidence level of $1-\alpha$ is specified, the goal is to construct a prediction interval $\hat C_{ij}(\alpha,\cO(X,M), M)$ such that
$\PP_{\cX}\sbr{X_{ij}\in \hat C_{ij}(\alpha,\cO(X,M), M)\given M}\geq 1-\alpha,\ (i,j)\in \Omega^c.$
As a result, by marginalizing over $M$,
\begin{equation}
    \PP\sbr{X_{ij}\in \hat C_{ij}(\alpha,\cO(X,M), M)}\geq 1-\alpha,\ (i,j)\in \Omega^c.
    \label{eq:coverage meaning}
\end{equation}
The second type of assumptions concerns the missingness mechanism. Assuming $M\sim\cM$, we then construct the prediction interval $\hat C_{ij}(\alpha,\cO(X,M), M)$ such that the coverage probability, conditional on $X$, is at least $1-\alpha$.
By marginalizing over $X$, we obtain the same result as in \eqref{eq:coverage meaning}.  For uncertainty quantification of the SIAP algorithm, we follow the second approach. We impose suitable assumptions on $\cM$ and construct prediction intervals using conformal prediction (CP).

\subsection{Conformal Prediction (CP) for Matrix Completion}
Conformal prediction (CP, \citealp{shafer_tutorial_2008}) is a Monte Carlo-based uncertainty quantification technique that does not rely on distributional assumptions but on the assumption of \textit{exchangeability}. The set of observed indices $\Omega$ is randomly divided into a training set $\cS_{\text{tr}}$ and a calibration set $\cS_{\text{cal}}$. 
Specifically, $\cS_{\text{tr}}=\cbr{(i,j)\in\Omega: S_{ij}=1}$ and $\cS_{\text{cal}}=\cbr{(i,j)\in\Omega: S_{ij}=0}$,
where $S_{ij}\overset{\text{i.i.d.}}{\sim} \text{Bern}(1-p_\text{cal})$. 
Define the absolute residual as $R_{ij}=|X_{ij}-\hat X_{ij}|$ for $i\in[m]\text{ and }\ j\in[n]$, where $\hat X_{ij}$ is estimated from the training data $\cP_{\cS_{\text{tr}}}(X)$.
$(R_1,\ldots,R_N)$ are \textit{exchangeable} if, given that they take the values $(r_1,\ldots,r_N)$ in no particular order, all $N!$ permutations are equally likely. 
\citet{gui_conformalized_2023} present prediction intervals under independent heterogeneous missingness. 
In the special case where each entry is missing independently with probability $p_0$, 
they construct a prediction interval $\hat C_{ij}(\alpha) = \hat X_{ij} \pm \hat q(\alpha)$ that achieves \textit{average coverage probability} guarantee that $\EE_\cM[\operatorname{AvgCov}(\widehat{C} ; X, \Omega)] \geq 1-\alpha$, where
\[
\hat q(\alpha) = \text{quantile}_{1-\alpha} \rbr{\sum_{(i,j)\in \cS_{\text{cal}}} \frac{1}{n_{\text{cal}}+1}\cdot \delta_{R_{ij}} +\frac{1}{n_{\text{cal}}+1}\cdot \delta_{+\infty}},
\]
\[
\operatorname{AvgCov}(\widehat{C} ; X, \Omega) = \frac{1}{|\Omega^c|} \sum_{(i,j)\in\Omega^c} \1\cbr{X_{ij} \in \hat C_{ij}},
\]
and $\delta_t$ denotes a point mass at $t$.


\subsection{Adapting CP to Mixed Missingness Patterns}
In the SSI data, the entries are not missing independently due to a mixture of the downtime and scatter missingness mechanism, which makes the application of CP nontrivial. 
We impose Assumption~\ref{ass:cp missingness} on the missingness mechanism for downtime and scatter missingness.
Suppose we observe $\bw$ and $O$, the downtime and scattered missing indicators, respectively, where $\bw\in\RR^n$ and $O \in\RR^{m\times n}$. The $j$-th element of $\bw$ equals 1 if the $\bx_j$ contains only missing values, and 0 otherwise. The $(i,j)$-th element of $O$ is 1 if $X_{ij}$ is missing, and 0 otherwise. Then the overall missingness indicator $M_{ij} = \mathrm{sign}(O_{ij}+w_{j})$, thereby defining a mixed missingness mechanism.

\begin{assumption}[A mixed, independent missingness mechanism]
    \normalfont
    \label{ass:cp missingness}
    \begin{align*}
     \text{Downtime}  \quad & P(w_{j}=1) = p,\ P(w_{j}=0) = 1-p,\text{ for }j\in[n]\text{ independently.}\\
       \text{Scattered}\quad &\cbr{O_{ij}}_{i\in[m],j\in[n]}\overset{\mathrm{i.i.d.}}\sim \mathrm{Bern}(p'). \\
       \text{Independence}\quad & O\perp \bw.
    \end{align*}
\end{assumption}

Since the missingness in $X$ arises from two independent mechanisms, the complete permutation exchangeability assumption proposed by \citet{gui_conformalized_2023} is no longer valid. Moreover, given the heterogeneous wavelength-dependent (i.e., row-wise) variance of the SSI matrix, we assume column-wise permutation exchangeability separately for downtime and scattered missingness, due to their independence.
Let $\hat X$ denote the estimated matrix. The following assumption formalizes the column permutation exchangeability.
\begin{assumption}[Column permutation exchangeability of residuals]
    \normalfont
    \label{ass:cp exchangeability}
    Suppose $\cS_{\mathrm{dt}}=\cbr{(i,j)|\bw_j=1}$, $\cS_{\mathrm{sc}}=\cbr{(i,j)|O_{ij}=1}$.
    We assume that given any $i\in[m]$,
    \[
    \begin{aligned}
        &\PP\sbr{(R_{i,j_1},\ldots, R_{i,j_{N_i^{\mathrm{dt}}}})=(r_1,\ldots,r_{N_i^{\mathrm{dt}}})\given \cbr{R_{i,j_k}}_{k=1}^{N_i^{\mathrm{dt}}}=\cbr{r_k}_{k=1}^{N_i^{\mathrm{dt}}}} \\
        &= \PP\sbr{(R_{i,\pi(j_1)},\ldots, R_{i,\pi(j_{N_i^{\mathrm{dt}}})})=(r_1,\ldots,r_{N_i^{\mathrm{dt}}})\given \cbr{R_{i,j_k}}_{k=1}^{N_i^{\mathrm{dt}}}=\cbr{r_k}_{k=1}^{N_i^{\mathrm{dt}}}},
    \end{aligned}
    \]
    for any ${(r_1,\ldots,r_{N_i^{\mathrm{dt}}}) \in \RR^{N_i^{\mathrm{dt}}}}$ where $N_i^{\mathrm{dt}}=\#\cbr{(i,j)|(i,j)\in\cS_{\mathrm{dt}},j=1,\ldots,n}$. Similarly, the same statement holds when $N_i^{\mathrm{dt}}$ and $\cS_{\mathrm{dt}}$ are replaced with $N_i^{\mathrm{sc}}$ and $\cS_{\mathrm{sc}}$.
\end{assumption}

Given these assumptions, we split $\Omega$ into a training set $\mathcal S_\text{tr}$, a downtime calibration set $\mathcal S_\text{cal1}$, and a scattered calibration set $\mathcal S_\text{cal2}$ as follows. 
First, for each $j\in \cC_{\text{dt}}^c$, where $\cC_{\text{dt}}\subset[n]$ denotes the set of downtime columns, we sample  $U_{j}\overset{\text{i.i.d.}}{\sim}\text{Bern}(p_\text{cal1})$ and define$\cS_{\mathrm{cal1}}=\cbr{(i, j) \in \Omega: U_{j}=1}$.
Next, on the remaining set $\Omega-\cS_{\mathrm{cal1}}$, we sample $\tilde U_{ij}\overset{\text{i.i.d.}}{\sim}\text{Bern}(p_\text{cal2}),\ (i,j)\in \Omega-\cS_{\mathrm{cal1}}$, and define
$\cS_{\mathrm{cal2}}=\cbr{(i, j) \in \Omega: \tilde U_{ij}=1}$.
The uncertainty intervals are then given by
\begin{equation}
        \hat C_{ij}(\alpha) = \left\{\begin{array}{cc}
        \hat X_{ij} \pm \hat q_i^{\mathrm sc}(\alpha), & \text{ if }(i,j)\text{ is a scattered missing entry, } \\
        \hat X_{ij} \pm \hat q_i^{\mathrm dt}(\alpha), & \text{ if }(i,j)\text{ is a downtime missing entry, } \\
    \end{array}\right.
    \label{eq:cp prediction interval}
\end{equation}
where 
\[
\hat q_i^{\mathrm sc}(\alpha) = \text{quantile}_{1-\alpha} \rbr{\sum_{j:(i,j)\in \cS_{\text{cal2}}} \frac{\delta_{R_{ij}}}{n^{\mathrm{cal2}}_i+1} +\frac{\delta_{+\infty}}{n^{\mathrm{cal2}}_i+1}},\ n^{\mathrm{cal2}}_i = \#\cbr{j:(i,j)\in\cS_\mathrm{cal2}},
\]
\[
\hat q_i^{\mathrm dt}(\alpha) = \text{quantile}_{1-\alpha} \rbr{\sum_{j:(i,j)\in \cS_{\text{cal1}}} \frac{\delta_{R_{ij}}}{n^{\mathrm{cal1}}_i+1}  +\frac{\delta_{+\infty}}{n^{\mathrm{cal1}}_i+1} },\ n^{\mathrm{cal1}}_i = \#\cbr{j:(i,j)\in\cS_\mathrm{cal1}}.
\]
We provide an illustration of the sample-splitting procedure and the associated exchangeability properties in the supplementary material (Figure S1).
Under Assumptions~\ref{ass:cp missingness} and~\ref{ass:cp exchangeability}, as a direct extension of \citet{gui_conformalized_2023}, the prediction intervals are guaranteed to achieve an expected \textit{global average coverage rate} of at least $1-\alpha$, i.e.,
$\EE_\cM[\operatorname{AvgCov}(\widehat{C} ; X, \Omega)] \geq 1-\alpha$.
Moreover, by using row-specific calibration sets, the \textit{local coverage rates} are also guaranteed, i.e.,
$\EE_\cM[\operatorname{AvgCov}_i(\widehat{C} ; X, \Omega)] \geq 1-\alpha,$
where 
\begin{equation}
    \operatorname{AvgCov}_i(\widehat{C} ; X, \Omega) = \frac{1}{|\cbr{j:(i,j)\in\Omega^c}|} \sum_{\cbr{j:(i,j)\in\Omega^c}} \1\cbr{X_{ij} \in \hat C_{ij}}.
    \label{eq:local coverage rates}
\end{equation}
In Section~\ref{sec:simulation}, we demonstrate via simulation studies that both the local and global coverage rates achieve a near-nominal level.

\section{SSI Imputation with Synthetic Missingness}
\label{sec:simulation}

In this section, we use a synthetic study to examine how several imputation methods behave under missingness patterns motivated by the SSI application. In Section~\ref{sec:related works}, we reviewed three types of modeling strategies that represent distinct ways of learning missing values from observed data: SIAP, TRMF and LATC as loss-based matrix completion methods with regularization, scalable GP as a Bayesian nonparametric smoother, and MARSS as a likelihood-based time series model. The purpose is not only to compare numerical performance but also to understand which features of the data are most consequential for a successful reconstruction. This is important for SSI because the data exhibit several unique characteristics that can be favorable to some methods and challenging to others: a pronounced periodic mean structure, substantial dependence across spectra, and missingness that may appear either as extended downtime or as scattered loss of observations. We compare these methods in terms of imputation accuracy, the reliability of uncertainty quantification, and computational efficiency. We also conduct a component-wise comparison across several reduced SIAP specifications to assess how individual characteristics of SSI contribute to imputation accuracy.

The complete synthetic SSI data from which we simulate missingness in this section is generated by averaging the empirical solar irradiance model NRLTSI2-NRLSSI2 and the semiempirical model SATIRE \citep{matthes_solar_2017} between March 14, 2018, and January 29, 2023, resulting in a $2104 \times 1783$ matrix with no missing entries. Missingness is then artificially imposed in two ways. First, columns are dropped independently with probability $p_0$ to simulate downtime. Second, among the remaining observed entries, an additional $100p_0\%$ are removed uniformly at random to mimic scattered missingness. We consider $p_0 \in \{0.1, 0.3, 0.5\}$ and repeat each setting 100 times. To obtain the uncertainty quantification for SIAP and its variants, the observed entries are further divided into training and conformal calibration sets using the sample splitting procedure described in Section~\ref{sec:cp}, with 90\% used for training and 10\% reserved for conformal calibration. 

The scalable GP model is implemented using the \texttt{GpGp} package in \texttt{R} \citep{guinness_permutation_2018}, with spline basis functions in time as covariates and an exponential isotropic kernel. The MARSS model is implemented using the \texttt{MARSS} package in \texttt{R} \citep{holmes_analysis_2024, holmes_marss_2024}. For computational feasibility, the data matrix is partitioned into blocks of size $10 \times n$, and the MARSS model is fitted independently on each block. Uncertainty for MARSS is quantified using the parametric bootstrap described in \citet{holmes_marss_2024}. For GP and MARSS, the training data are preprocessed by temporal first-order differencing and row-wise standardization so that each wavelength series has zero mean, unit variance, and improved stationarity. 
For all SIAP-type models, we set the maximum rank to $r=10$ in both stages, $\lambda=5$, $\lambda_1=\lambda_2=3$, $\alpha=3$, maximum lag $p=2$, and the rank of $LL^\top$ in the cross-sectional covariance model to 100 when applicable. To improve numerical stability, row-wise standardization is performed. Batch experiments are managed using the \texttt{R} package \texttt{batchtools} \citep{lang_batchtools_2017}.

\begin{figure}[tb]
    \centering
    \includegraphics[width=0.8\linewidth]{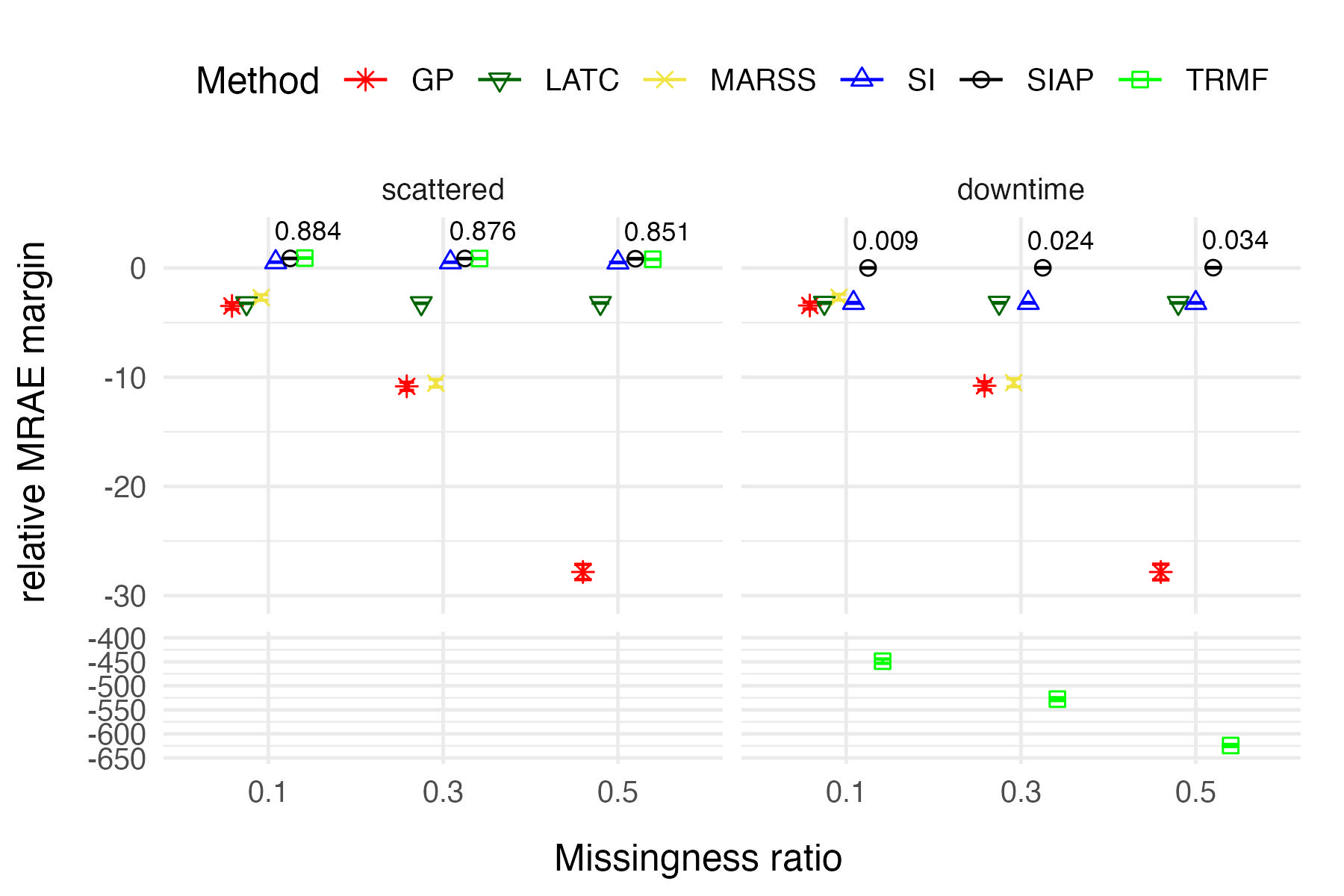}
    \caption{Imputation accuracy of selected methods with varying missingness ratio. The scatter points are the mean of the test set relative MRAE margin and the error bars give the 95\% confidence intervals. The MRAE margin is calculated relative to periodic cubic spline fitting.}
    \label{fig:simulation rel mrae margin combined}
\end{figure}

\begin{figure}[tb]
    \centering
    \includegraphics[width=0.9\linewidth]{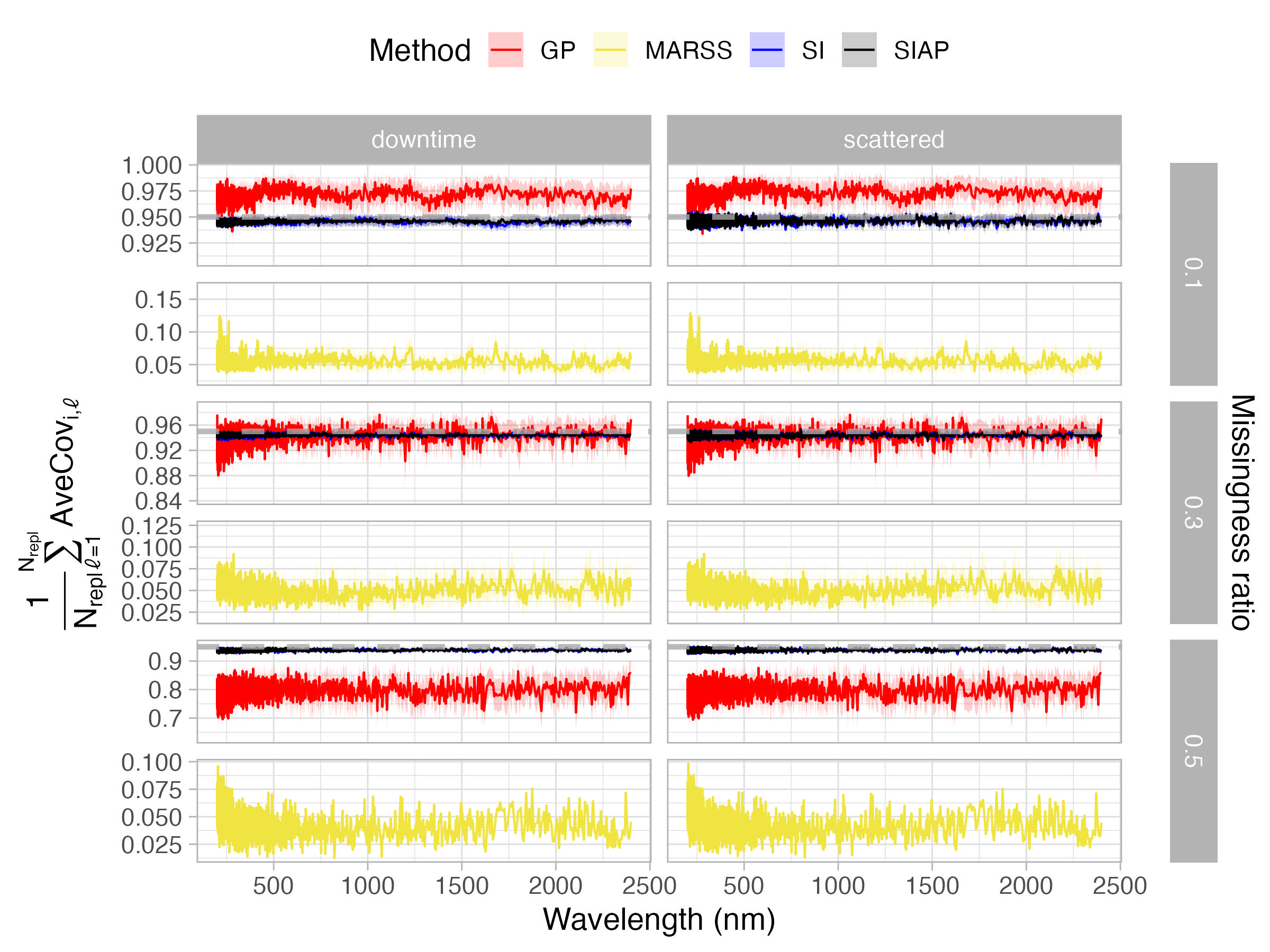}
    \caption{The average coverage rate, $\frac{1}{N_{\text{repl}}} \sum_{\ell=1}^{N_{\text{repl}}} \text{AveCov}_{i,\ell}$, where $\text{AveCov}_{i,\ell}$ is the \emph{local coverage rates} at the $i$-th row (see Eq~\ref{eq:local coverage rates}) in the $\ell$-th replicate. The shaded areas represent the 95\% empirical confidence interval computed over $N_{\text{repl}}=100$ replicates.}
    \label{fig:coverage vs wvl}
\end{figure}

We first evaluate the imputation accuracy using the entry-wise relative absolute error and the mean relative absolute error over an index set $\mathcal S \subset [m]\times[n]$,
\[
\mathrm{RAE}_{ij} = \frac{|\hat X_{ij} - X_{ij}|}{|X_{ij}|}\quad \text{and}\quad  \mathrm{MRAE}_{\mathcal S}
=
\frac{1}{|\mathcal S|}
\sum_{(i,j)\in \mathcal S}
\frac{|\hat X_{ij} - X_{ij}|}{|X_{ij}|}.
\]
To facilitate comparison across methods, we also report the relative MRAE margin with respect to a baseline model $\mathcal M_0$:
\[
\delta \mathrm{MRAE}_{\mathcal S}^{\mathcal M}
=
\frac{
\mathrm{MRAE}_{\mathcal S}^{\mathcal M_0}
-
\mathrm{MRAE}_{\mathcal S}^{\mathcal M}
}{
\mathrm{MRAE}_{\mathcal S}^{\mathcal M_0}
}.
\]
Intuitively, $\delta \mathrm{MRAE}_{\mathcal S}^{\mathcal M}$ quantifies how much better model $\cM$ outperforms the baseline $\cM_0$ on $\cS$. In the numerical study, we consider $\cM_0$ to be the periodic cubic splines using evenly spaced internal knots. Let $\mathcal S_w$ and $\mathcal S_o$ denote the sets of the downtime and scattered missingness. Figures~\ref{fig:simulation rel mrae margin combined} summarizes $\delta \mathrm{MRAE}_{\mathcal S_w}$ and $\delta \mathrm{MRAE}_{\mathcal S_o}$ for each method, from which we summarize our finds as follows. First, among the matrix-completion-based methods (SIAP, TRMF and LATC), SIAP gives the strongest overall performance and remains stable as the missingness ratio increases. It improves the imputation accuracy on TRMF and LATC, suggesting that explicitly accounting for the periodic mean structure is important in this application. 
Second, the smooth model-based methods (GP and MARSS) can also achieve good imputation accuracy. In additional runs without row-wise first-order differencing (which was used to make the time series zero-mean and stationary), and using only row-wise standardization, GP and MARSS can produce point estimates that are even more accurate than those from the matrix-completion-based methods (see Figure S2 in the supplement). This is not entirely surprising for SSI, which is generally stable over time and often exhibits limited abrupt local variation, so that interpolation-based and smoothing-based procedures can be effective. This suggests that GP and MARSS are more sensitive to preprocessing choices. Meanwhile, the model's good performance is subject to the Gaussian assumption, which may not be satisfied by the real data. GP is commonly considered robust to linear transformations, which means that taking the first-order difference should not alter Gaussianity. However, Figure~\ref{fig:coverage vs wvl} shows that GP intervals fail to achieve the nominal coverage, indicating that the data may not satisfy the Gaussian assumption anyway. Finally, learning favorable kernel parameters in GP requires additional effort. For example, when it is given data evenly scattered around zero, the scale parameter in the squared exponential kernel strongly influences whether GP treats the data as pure noise (with a large scale parameter) or as an informative curve hovering around the x-axis (with a small scale parameter). In our synthetic data imputation setting, GP is expected to treat the data as informative. However, because a large scale parameter in the GP kernel also produces a high marginal likelihood, the optimization process may converge to that solution. This is a major reason why GP yields worse imputation with row-wise differencing than without it.


Next, we conduct a detailed component-wise analysis of SIAP to illustrate how different charateristic of the SSI data is handled for effectively in SIAP. We compare several reduced SIAP specifications corresponding to the modules in Figure~\ref{fig:siap modules}: (1) SoftImpute (\texttt{SI}), which applies only module 2, and therefore uses only the low-rank information; (2) \texttt{SI PrdMu}, which combines modules 1 and 2, only imposing the periodic mean structure; (3) \texttt{SI PrdMu CrosSpec} (equivalent to Step~1 of SIAP), which combines modules 1--3, including both periodic mean and cross-spectral dependence, but does not separate downtime and scattered missingness; (4) \texttt{SIA}, which uses only module 5, including only the temporal smoothing (without periodic mean structure); (5) \texttt{SIA LatentPrdMu} (equivalent to Step~2 of SIAP), which combines modules 4 and 5, considering both the temporal smoothing and periodic mean; (6) \texttt{Step~1 + SI LatentPrdMu}, which combines module 1--4, incorporating all the components but temporal smoothing; and (7) the full SIAP model.

\begin{figure}[tb]
    \centering
    \includegraphics[width = 0.9\linewidth]{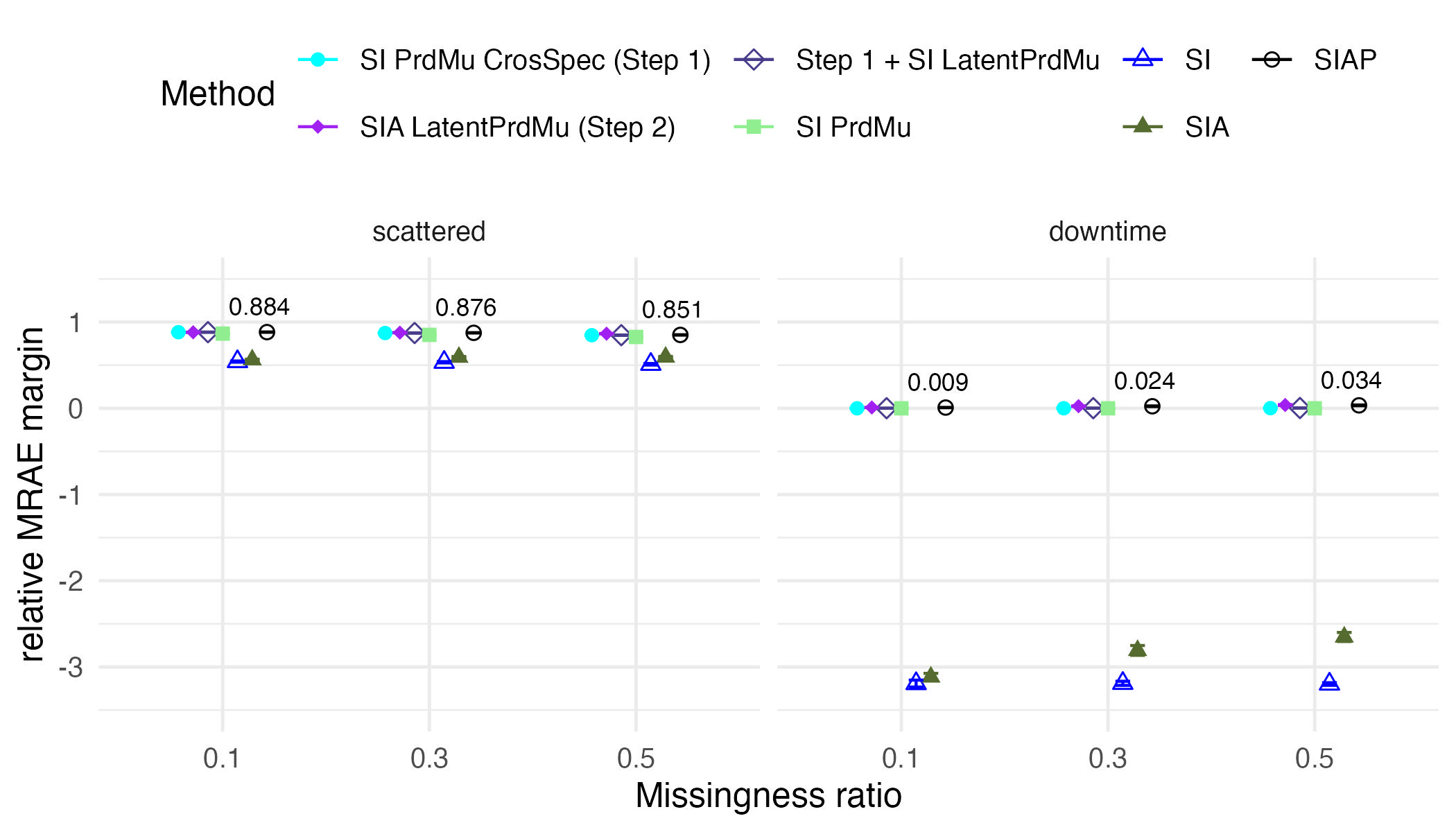}
    \caption{Component-wise study of SIAP. The scatter points are the mean of the test set relative MRAE margin and the error bars give the 95\% confidence intervals. }
    \label{fig:component wise}
\end{figure}

First, detrending is essential. Methods that omit the periodic mean adjustment, such as \texttt{SI} and \texttt{SIA}, perform substantially worse than their detrended counterparts. Second, modeling cross-sectional covariance improves the imputation of scattered missing entries, which in turn leads to better performance for downtime imputation. This can be seen by comparing \texttt{SI PrdMu} with \texttt{SI PrdMu CrosSpec} in Figures~\ref{fig:component wise}. Third, AR regularization in the latent temporal factors is particularly helpful for downtime recovery. When the scattered component has already been well reconstructed, incorporating this temporal regularization yields further gains over specifications without it, as illustrated by the comparison between \texttt{Step~1 + SI LatentPrdMu} and the full \texttt{SIAP}. Therefore, the advantage of \texttt{SIAP} does not come from a single modeling choice, but from combining periodic detrending, borrowing of strength across wavelengths, and temporal stabilization over longer gaps.

We next compare uncertainty quantification across methods. Figure~\ref{fig:coverage vs wvl} shows the average coverage rate by wavelength, denoted $\operatorname{AvgCov}_i$, with averages taken over the 100 replicates. The conformal intervals used with SIAP remain close to the nominal 95\% level across wavelengths and across all missingness regimes. Notice that the empirical coverage is slightly below nominal, at roughly 94\%, which is plausible because the actual SSI missingness mechanism is dependent and therefore does not exactly satisfy the exchangeability condition underlying conformal calibration. Even so, the conformal procedure remains stable in this setting. In particular, using column permutation exchangeability yields both reliable overall coverage and good wavelength-specific coverage. 
The GP intervals perform reasonably well when the missingness ratio is low, but their coverage deteriorates when 50\% of entries are missing. 
The MARSS intervals, constructed through parametric bootstrap under an i.i.d.\ Gaussian noise assumption, exhibit poor coverage throughout. These results suggest that strong distributional assumptions may not be the main obstacle for point prediction on SSI, but they become much more consequential when uncertainty quantification is required. 
Figure S3 in the supplementary material shows that the interval lengths of SIAP are comparable to those of GP. All intervals considered here are pointwise rather than simultaneous.

Runtime provides an additional practical comparison. In our experiments, scalable GP is the fastest method. SIAP has a comparable runtime, whereas MARSS is the slowest and it has to be fitted blockwise. 
This reflects an application tradeoff. Scalable GP offers computational efficiency and, when preprocessing is favorable, can provide very accurate point estimates. The SIAP-CP method is less dependent on such choices and gives a more stable combination of point accuracy and interval coverage across missingness settings. The primary computational cost of SIAP lies in memory allocation rather than the EM-based optimization, which typically converges in fewer than 10 iterations. MARSS is more computationally demanding and, in this application, does not offer a comparable gain in uncertainty quantification.

Overall, the synthetic study highlights several features of SSI reconstruction. Because SSI is stable, smooth model-based approaches such as GP and MARSS can produce good point estimates when the preprocessing is well selected. At the same time, their uncertainty quantification is more sensitive to modeling assumptions and becomes less reliable as missingness increases. However, reliable uncertainty quantification requires weaker and more transparent assumptions. 
On this dataset, SIAP provides a favorable compromise among accuracy, interval calibration, and runtime while aligning well with the structural features of SSI.
The componentwise comparisons further indicate that this performance is driven by three data-dependent ingredients: detrending to capture periodic mean structure, cross-sectional borrowing across wavelengths, and temporal regularization to stabilize recovery over longer gaps. Through this synthetic study, we not only present a benchmark, but also provide a way to identify which aspects of the SSI data are most consequential for successful imputation.

\section{Analysis of SSI Reconstruction} 
\label{sec:realdata}

In this section, we analyze daily TSIS-1 spectral solar irradiance (SSI) observations from March 14, 2018 to January 29, 2023, recorded on 2104 wavelength channels \citep{richard_tsis_2025}. We still compare the three types of imputation methods as we did in Section~\ref{sec:simulation}, together with evaluations of scientifically meaningful summary quantities derived from SSI observations. We validate our findings in Section~\ref{sec:simulation} again with this partially observed data.

As a fundamental \emph{climate data record}, SSI reconstruction supports downstream questions about solar variability and climate change. The relevant signals of interest include solar-cycle modulation, longer-term trends, and solar-cycle–driven changes such as atmospheric chemistry and radiative forcing, which are based on quantities derives from TSIS-1 SSI.
Therefore, we need to control both bias and variance for the SSI reconstruction, since small errors at the spectral level can propagate into substantial ones in quantities such as band-integrated irradiance and long-term trends. The National Polar-Orbiting Operational Environmental Satellite System (NPOESS) program established the measurement requirements \citep[Table 1]{mineart_national_2008, coddington_solar_2016} driven by the need to understand Earth’s climate response to solar radiation variability.
Let $I(\lambda,t)$ denote SSI at wavelength $\lambda$ and time $t$, the Earth's response to solar radiation are typically expressed through functionals such as band-integrated irradiance $B_{\Lambda}(t)=\int_{\Lambda} I(\lambda,t)\,d\lambda$, anomalies $I(\lambda,t)-\bar I(\lambda)$, and trend estimates from simple models like $B_{\Lambda}(t)=\alpha_{\Lambda}+\beta_{\Lambda} t+\varepsilon_t$. Writing $\widehat I(\lambda,t)=I(\lambda,t)+b(\lambda,t)+\delta(\lambda,t)$, the systematic bias $b(\lambda,t)$ governs absolute accuracy (quantified by the percentage relative absolute error) and the random error $\delta(\lambda,t)$ governs precision and propagates into uncertainty for derived quantities $\hat B_{\Lambda}$ and $\hat \beta_\Lambda$ via standard variance propagation.

Motivated by the need to detect small SSI signals over decades, published SSI CDR requirements therefore specify stringent performance targets with $0.2\%$ absolute accuracy, $0.01\%$ relative precision \citep{coddington_climate_2015, coddington_solar_2016}, and in TSIS-1 Level-1 requirements, an explicit $1$-sigma uncertainty goal of 0.2\% (maximum allowable threshold 1\%) \citep{richard_si-traceable_2020}. 
Therefore, based on these accuracy and uncertainty requirements, the analysis of the TSIS-1 data is designed to evaluate not only whether a method accurately fills in missing values, but also whether it provides uncertainty statements that remain reliable. We compare the reconstructed values with TSIS-1 observations and with SSI observations from another independent source, CSIM Level 3 photodiode SSI data \citep{laboratoryforatmosphericandspacephysicslasp_compact_2019} in order to assess both imputation accuracy and the reliability of the interval estimates. We tune SIAP’s hyperparameters sequentially; additional details are provided in the supplementary material.

First, to evaluate accuracy, we randomly hold out \(10\%\) of the observed columns and \(10\%\) of the remaining observed entries and treat them as test sets representing downtime and scattered missingness, respectively. For interval estimation with conformal prediction, the remaining observed entries are further split into training and calibration sets. We summarize point prediction performance using entry-wise relative absolute error and evaluate interval performance through empirical coverage in Table~\ref{tab:realdata mrae margin to mean imputation} and Table~\ref{tab:realdata coverage}. 
SIAP remains accurate across both scattered and downtime missingness because it exploits the features that are directly relevant to this dataset, namely cross-spectral structure, a periodic mean component, and independent downtime and scattered missingness. More importantly, the conformal intervals achieve coverage that is substantially more reliable than GP or MARSS intervals, whose validity depends on distributional assumptions that appear to be strained in this application. The under-coverage is especially pronounced in high-variability wavelengths such as those around 400nm. 
This reveals an important limitation of model-based inference (e.g., GP and MARSS): once stronger structural assumptions are imposed to stabilize fitting under the observed missingness patterns, point accuracy may deteriorate, and interval calibration can become unreliable. 

\begin{table}[t]
    \centering
    \caption{Summary of entry-wise RAE margins w.r.t. imputation by global mean value.}
    \label{tab:realdata mrae margin to mean imputation}
    \begin{tabular}{lcccccccc}
        \toprule
        Type 
        & \multicolumn{2}{c}{SIAP (\( \times 10^{-3} \))} 
        & \multicolumn{2}{c}{GP (\( \times 10^{-3} \))} 
        & \multicolumn{2}{c}{MARSS (\( \times 10^{-3} \))} 
        & \multicolumn{2}{c}{SI (\( \times 10^{-3} \))} \\
        \cmidrule(lr){2-3} \cmidrule(lr){4-5} \cmidrule(lr){6-7} \cmidrule(lr){8-9}
        & Mean & SE & Mean & SE & Mean & SE & Mean & SE \\
        \midrule
        Scattered & \textbf{2.42} & 0.0084 & -6.54 & 0.0215 & -4.30 & 0.0191 & 2.41 & 0.0084 \\
        Downtime & \textbf{2.07} & 0.0062 & -6.58 & 0.0182 & -4.28 & 0.0158 & 0.0016 & 0.0002 \\
        \bottomrule
    \end{tabular}
\end{table}

\begin{table}[t]
    \centering
    \caption{Average coverage rate calculated on the test set.}
    \begin{tabular}{c|cll}
         \toprule
         &   SIAP & GP&MARSS\\
         \midrule
         overall& 0.953& 0.931&0.076\\
         downtime& 0.958& 0.930&0.075\\
         scattered& 0.946& 0.932&0.076\\
         \bottomrule
     \end{tabular}
    \label{tab:realdata coverage}
\end{table}

\begin{figure}[tb]
    \centering
    \includegraphics[width=\linewidth]{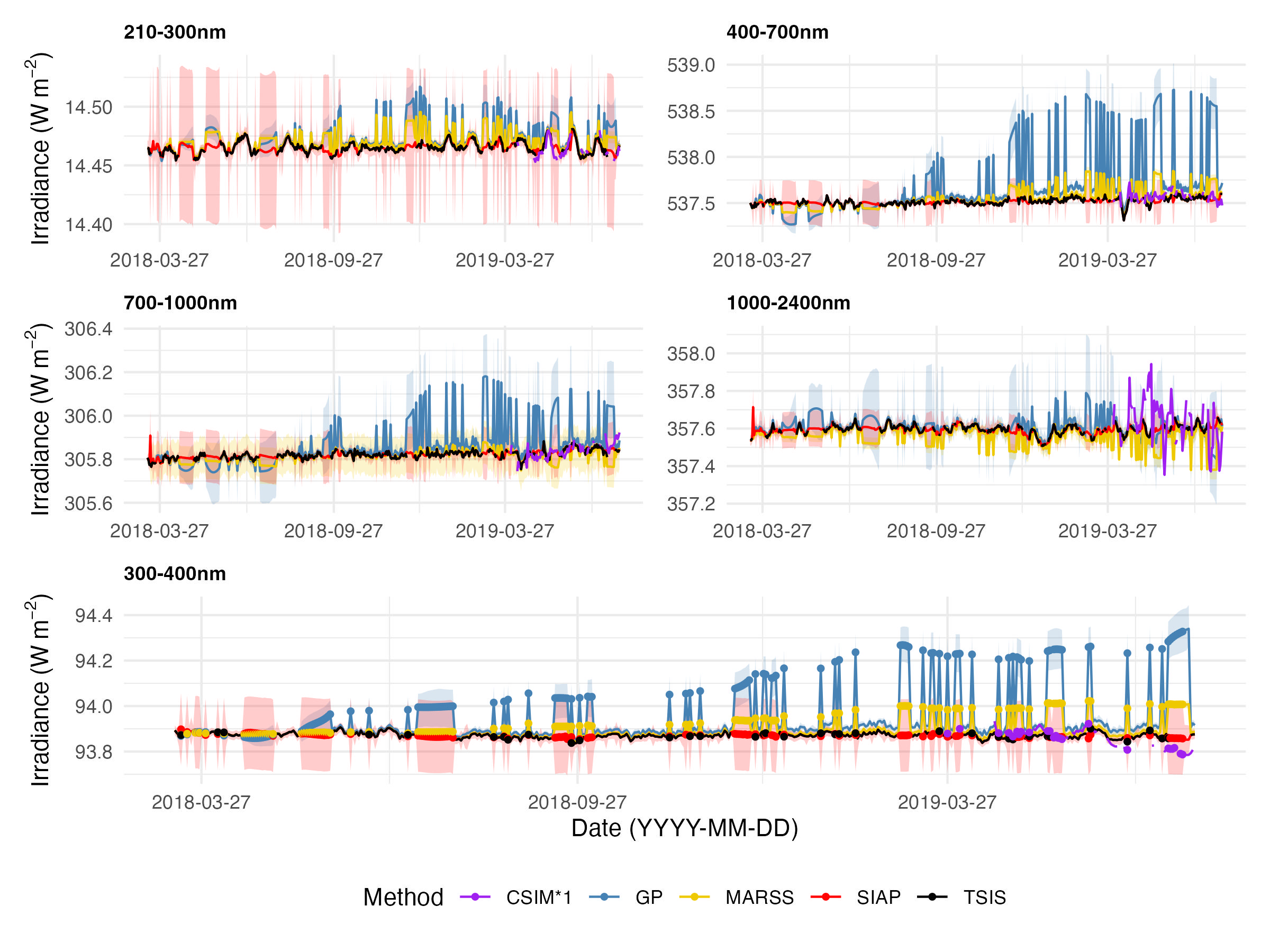}
    \caption{Integrated SSI imputation results and CSIM observations across five wavelength bands spanning the ultraviolet to near-infrared. The 300–400 nm band is shown enlarged, with solid points marking the downtime missingness. CSIM SSI values are scaled so that measured and imputed irradiances agree on a reference day in June 2019. TSIS-1 covers 2104 spectral channels from 200.015 nm to 2399.011 nm; CSIM covers 2343 channels from 210.014 nm to 2596.299 nm. The two datasets share a partial temporal overlap.}
    \label{fig:realdata binned SSI}
\end{figure}

Next, we examine the reconstructed series at a scientifically important scale by integrating SSI over 210--300\,nm 300--400\,nm, 400--700\,nm, 700--1000\,nm and 1000--2400\,nm bands, among which the temporal variability at the 300--400\,nm band is among the largest across the spectrum. 
Figure~\ref{fig:realdata binned SSI} compares the reconstructed band-integrated irradiance with TSIS-1 observations and with CSIM Level 3 photodiode SSI data \citep{laboratoryforatmosphericandspacephysicslasp_compact_2019}. Because CSIM and TSIS-1 are produced by different instruments and exhibit systematic measurement differences, this comparison is used as a qualitative external check. Accordingly, the CSIM series is scaled to match TSIS-1 on a reference day in June 2019. 
GP tends to over-smooth when the scale parameter is large, so imputed values inherit the sign of neighboring observed values, as visible in Figure~S5 in the supplement. After back-transformation, this behavior produces the upward-trending GP curve. This also suggests that the downtime missingness mechanism may not be missing at random in practice, since otherwise the GP curve would go up and down around the observed curve, rather than exhibiting a consistent upward trend, as can be seen in the 300--400\,nm band subfigure in Figure~\ref{fig:realdata binned SSI}. Table~\ref{tab:downtime missingess type summary} summarizes the distribution of chunk types across wavelength bands, where each chunk is a maximal contiguous block of missing time points classified by the sign of its two flanking observed values in the preprocessed TSIS-1 data: two positive ($(+,+)$), two negative ($(-,-)$), or mixed-sign neighbors ($(+,-)$). Instead, in the 300--400\,nm, 400--700\,nm and 700--1000\,nm bands, there are more $(+,+)$ chunks than $(-,-)$ chunks. The weighted proportions, which upweight longer chunks, tell a similar story. This is consistent with Figure~\ref{fig:realdata binned SSI}, where the GP imputations for 300--400\,nm, 400--700\,nm and 700--1000\,nm bands show a clear upward trend.
Looking at the uncertainty bands, especially where variability is strongest,  the conformal intervals built on SIAP are better aligned with the observed fluctuations, whereas GP and MARSS often produce uncertainty bands that are too narrow to accommodate the realized TSIS-1 values. Taken together, smooth model-based imputation may look attractive for point estimation on stable SSI series, yet assumption-lean imputation and uncertainty quantification is crucial when the scientific goal is a climate-facing reconstruction whose uncertainty must remain interpretable in downstream use.

\begin{table}[t]
    \centering
    \caption{Summary of downtime missingness chunk types in band-integrated SSI, classified by the sign of their flanking observed values. Weighted proportions are computed by weighting each chunk by its width (number of missing time points). Total number of chunks: $n_{\text{chunks}} = 176$.}
    \begin{tabular}{lcccccc}
        \toprule
        $n_{\text{chunks}} = 176$ & \multicolumn{3}{c}{Unweighted} & \multicolumn{3}{c}{Width-weighted} \\
        \cmidrule(lr){2-4} \cmidrule(lr){5-7}
        Band & $(+,+)$ & $(-,-)$ & $(+,-)$ & $(+,+)$ & $(-,-)$ & $(+,-)$ \\
        \midrule
        210--300nm    & 0.2273 & 0.2727 & 0.5000 & 0.1987 & 0.2301 & 0.5712\\
        300--400nm    & \textbf{0.3182} & \textbf{0.2045} & 0.4773 & \textbf{0.3361} & \textbf{0.1705} & 0.4934\\
        400--700nm    & \textbf{0.2727} & \textbf{0.2045} & 0.5227 & \textbf{0.3096} & \textbf{0.1921} & 0.4983\\
        700--1000nm   & \textbf{0.2500} & \textbf{0.1648} & 0.5852 & \textbf{0.2732} & \textbf{0.1623} & 0.5646\\
        1000--2400nm  & 0.2102 & 0.2273 &0.5625 & 0.1821 &	0.2201987	\\
        \bottomrule
    \end{tabular}
    \label{tab:downtime missingess type summary}
\end{table}

The main message of this analysis is therefore twofold. First, the TSIS-1 SSI data provide a valuable applied setting in which different statistical reconstruction strategies behave differently for scientifically interpretable reasons. Second, in this setting, point imputation performance and uncertainty calibration should be assessed separately: smoothing-based models can be competitive for point prediction, while distribution-free intervals can be more reliable when model assumptions are difficult to justify.

\section{Conclusion}
\label{sec:conclusion}

In this work, we address the problem of reconstructing missing values in the TSIS-1 SSI data with quantified uncertainties. Unlike methods that rely on external solar proxies or physics-based semi-empirical modeling, our approach is purely data-driven and leverages only the observed SSI itself. This makes our method particularly valuable in contexts where auxiliary solar indices or models are unavailable and/or unreliable.

Our imputation strategy is built on matrix completion techniques, modified based on domain-specific knowledge such as the temporal periodicity of solar activity and the strong spectral correlations across wavelength channels. While classical matrix completion methods often assume missing uniformly at random and linear stationarity, these assumptions are violated in SSI data due to systematic downtime missingness and non-stationary solar cycle progression. Accordingly, we incorporated smoothness penalties and periodic components into the matrix factorization framework. This allows us to flexibly capture both long-term solar trends and short-term fluctuations without overfitting.

Experimental results demonstrate that our methods have competitive performance compared to standard imputation techniques such as Gaussian process kriging and time series smoothing. Our proposed method is capable of handling entire missing columns (i.e., downtime), a task that conventional matrix completion methods struggle with. The application of conformal prediction equips our method with prediction intervals for each entry, enabling risk-aware downstream analyses and physical model validations. 

In conclusion, this study presents a flexible, interpretable, and self-contained method for SSI imputation that avoids reliance on external covariates. We also provide a systematic comparison of the strengths and weaknesses of three major types of matrix completion algorithms (loss-function-based matrix factorization, likelihood-based parametric time series, and Bayesian non-parametric Gaussian process models), using synthetic simulations and real data. By addressing the unique challenges posed by the SSI data --- including non-uniformly-at-random missingness and periodic dynamics --- our proposed method contribute a robust and scalable tool for the reconstruction and usability of SSI records in climate science, atmospheric modeling, and solar physics.  Our method is also generalizable and could be adapted to other geophysical datasets with similar structural and temporal characteristics.

\begin{acks}[Acknowledgments]


\end{acks}

\begin{funding}
Y. Ke, X. Huang, and O. Coddington are supported by NASA’s Living With a Star program under grant 80NSSC21K1306 awarded to the University of Michigan, with a subcontract to the University of Colorado Boulder. 
\end{funding}

\begin{supplement}
\stitle{Supplementary Material to ``Matrix Factorization-Based Solar Spectral Irradiance Missing Data Imputation with Uncertainty Quantification''}
\sdescription{Proofs of convergence theorems and supplementary experimental results.}

\end{supplement}

\begin{supplement}
\stitle{Data and source code.}
\sdescription{The \href{https://deepblue.lib.umich.edu/data/concern/data_sets/5d86p133v}{synthetic benchmark data} and \href{https://deepblue.lib.umich.edu/data/concern/data_sets/rx913r011}{reconstructed SSI data product} are publicly available, together with open-source code at the \href{https://github.com/walnut-7/siap.git}{GitHub repository}, to support reproducibility and facilitate further scientific use.}

\end{supplement}


\bibliographystyle{template_aoas/imsart-nameyear} 
\bibliography{ref}       


\end{document}


\begin{frontmatter}
\title{Supplementary Material to ``Matrix Factorization-Based Solar Spectral Irradiance Missing Data Imputation with Uncertainty Quantification''}

\begin{aug}
\author[A]{\fnms{Yuxuan}~\snm{Ke}
    \ead[label=e1]{yuxuank@umich.edu}
    \orcid{0009-0003-1119-0807}
    },
\author[B]{\fnms{Xianglei}~\snm{Huang}
    \ead[label=e2]{xianglei@umich.edu}
    \orcid{0000-0002-7129-614X}
    },
\author[C]{\fnms{Odele}~\snm{Coddington}
    \ead[label=e3]{Odele.Coddington@lasp.colorado.edu}
    \orcid{}
    }
\and
\author[A]{\fnms{Yang}~\snm{Chen}
    \ead[label=e4]{ychenang@umich.edu}
    \orcid{0000-0002-9516-8134}
    }

\address[A]{Department of Statistics,
University of Michigan \printead[presep={,\ }]{e1,e4}}

\address[B]{Department of Climate and Space Science, University of Michigan \printead[presep={,\ }]{e2}}

\address[C]{Laboratory For Atmospheric And Space Physics,      University of Colorado Boulder \printead[presep={,\ }]{e3}}

\end{aug}

\end{frontmatter}

\begin{table}[h]
    \centering
    \begin{tabular}{c|l}
        \toprule
        Notation & Description \\
        \midrule
        $Q^{n_1}_\cS$, where $\cS\subset\NN$. & $Q^{n_1}_\cS =\diag(q_1,\ldots,q_{n_1})$, where $q_j= \mathbbm 1\rbr{j\in\cS}$. \\
        \multirow{2}{*}{$\cP_\Omega(\cdot), \cP_\Omega^\perp(\cdot)$} & For any matrix $A\in \RR^{m\times n}$, $\cP_\Omega(A),\cP_\Omega^\perp(A)\in\RR^{m\times n}$ s.t. \\
         & $\cP_\Omega(A)_{ij}=\left\{\begin{array}{cc}
              A_{ij}, & (i,j)\in\Omega  \\
              0, & (i,j)\notin\Omega
         \end{array}\right.$, 
         $\cP_\Omega^\perp(A)_{ij}=\left\{\begin{array}{cc}
              A_{ij}, & (i,j)\notin\Omega  \\
              0, & (i,j)\in\Omega
         \end{array}\right.$. \\
         $P_n(k)$ & Permutation matrix. 
            $P_n(k)=\begin{pmatrix}
            {0}_{(n-k)\times k}& I_{n-k} \\
            I_k &{0}_{k\times(n-k)}
         \end{pmatrix},\ k=1,\cdots,n$. \\
         %
         %
         %
         $\otimes$ & Kronecker product of matrices. \\
         $\odot$ & Entry-wise multiplication. \\
        \bottomrule
    \end{tabular}
    \caption{Notations used in the supplementary material.}
    \label{tab:app notations}
\end{table}

\section{Connections between ECM and \texttt{SoftImpute-ALS}}
\label{app:ecm analogue}
The objective function of \texttt{SoftImpute-ALS} has a deep connection with the likelihood function under probability
\begin{equation}
    \vvec(X)\given A,B \sim \cN(\vvec(AB^{\top}), \sigma^2 I_{mn}),
    \label{eq:MF prob}
\end{equation}
where $\vvec$ denotes the vectorization of matrix $X$. The log-likelihood of $A$ and $B$ is
\[
\log p(X|A,B) = -\frac{1}{2\sigma^2}\fnorm{X-AB^{\top}}^2 - \frac{mn}{2}\log \sigma + const.
\]
Integrating out the missing entries in $X$ yields the marginal log-likelihood
\begin{equation}
    \log p(X_\cO|A,B) = 
    -\frac{1}{2\sigma^2}\fnorm{\cP_\Omega(X-AB^{\top})}^2 - \frac{|\Omega|}{2}\log \sigma + const,
    \label{eq:MF marginal likelihood}
\end{equation}
where $X_\cO$ and $X_\cM$ denote the observed and missing parts of $X$, respectively. Consider the Frobenius norm regularized maximum likelihood problem, embedding $\sigma$ within the tuning parameter:
\[
\begin{aligned}
    &\min_{A,B}\left\{ \log p(X_\cO|A,B) + \frac{\tilde\lambda}{2}(\fnorm{A}^2 + \fnorm{B}^2)\right\} \\
    \Leftrightarrow & \min_{A,B} \left\{\fnorm{\cP_\Omega(X-AB^{\top})}^2 + \lambda(\fnorm{A}^2 + \fnorm{B}^2)\right\}.
\end{aligned}
\]

Write $X=\cbr{X_\cO,X_\cM}$ and consider estimating $X_\cM$ using the expectation conditional maximization (ECM) algorithm. 
The regularized ECM update is the same as the \texttt{SoftImput-ALS} iteration. In the E-step, we take the expectation of the regularized log-likelihood defined as
\[
l_\lambda(X|A,B) = - \frac{1}{2}\fnorm{X-AB^{\top}}^2 - \frac{\lambda}{2}(\fnorm{A}^2 + \fnorm{B}^2).
\]
By taking the expectation of $l_\lambda$ and alternately maximizing over $A$ and $B$, we obtain the following updates. The E-step computes the expectated log-likelihood, yielding
\begin{align*}
    Q(A,B|A_k,B_k) &= \EE_{X_\cM|X_\cO,A_k,B_k}\sbr{l_\lambda(X|A,B))} \\
    &= - \frac{1}{2}\fnorm{\cP_\Omega(X)+\cP_\Omega^\perp(A_kB_k^{\top})-AB^{\top}}^2 - \frac{\lambda}{2}(\fnorm{A}^2 + \fnorm{B}^2).
\end{align*}
The M-steps update the $A$ and $B$ matrices by 
\begin{align}
    & A_{k+1}= \argmax_{A} Q(A,B_k|A_k,B_k),
    \label{eq:regularized M1} \\
   & B_{k+1}= \argmax_{B} Q(A_{k+1},B|A_{k+1},B_k).
    \label{eq:regularized M2}
\end{align}
Equations~\eqref{eq:regularized M1} and \eqref{eq:regularized M2} correspond exactly to the approach used by \citet{hastie_matrix_2015} to solve the \texttt{SoftImpute-ALS} problem.

\section{Derivation of SIAP update}

Table~\ref{tab:app notations} lists the notations used in this section.

\subsection{Step 1}
\label{app:1 deriv}



\subsubsection{E-step}

The E-step involves computing all the first- and second-order conditional expectations, which are 
\begin{align}
    \conde{\bx_{j,{\cM_j}}} &= \bmu_{j,{\cM_j}}^k + \Sigma_{j,{\cM_j}{\cO_j}}^k{(\Sigma_{j,{\cO_j}}^k)}^{-1}(\bx_{j,{\cO_j}} - \bmu_{j,{\cO_j}}^k), 
    \label{eq:<x>} \\
    \conde{\bx_{j,{\cM_j}}\bx_{j,{\cM_j}}^{\top}} &=\Sigma_{j,{\cM_j}}^k - \Sigma_{j,{\cM_j}{\cO_j}}^k{(\Sigma_{j,{\cO_j}}^k)}^{-1}\Sigma_{j,{\cO_j}{\cM_j}}^k+ \conde{\bx_{j,{\cM_j}}}\conde{\bx_{j,{\cM_j}}^{\top}},
    \label{eq:<xx>}
\end{align}
\begin{align}
    \conde{\bz_j} &= \Sigma_zL^{\top}\Lambda^{-1}\conde{\tilde\bx_j}, \label{eq:<z>}\\
    \conde{\bz_j \bz_j^{\top}} &= \Sigma_z + \Sigma_zL_k^{\top}\Lambda_k^{-1}\conde{\tilde \bx_j\tilde \bx_j^{\top}}\Lambda_k^{-1}L_k\Sigma_z, \label{eq:<zz>}\\
    \conde{\tilde \bx_j\bz_j^{\top}} &= \conde{\tilde \bx_j\tilde \bx_j^{\top}} \Lambda_k^{-1}L_k\Sigma_z, \label{eq:<xz>}
\end{align}
where $\tilde\bx_j=\bx_j-\bmu_j^k$, $\bmu_j^k=A_k\bb_j^k+\Theta_k^{\top}\bphi_j$, $\Sigma_z = (I+L_k^{\top}\Lambda_k L_k)^{-1}$, $\Sigma_k = \Lambda_k + L_kL_k^{\top}$. Derivations are provided in Appendix Section~\ref{app:1 deriv}. We omit these derivations for Eqs.~\eqref{eq:<x>} and~\eqref{eq:<xx>}, as they follow directly from standard results for conditional Gaussian distributions.

Let $\tilde \bx_j=\bx_j-(A\bb_j+\Theta^{\top}\bphi_j)$, $\bz_j\sim \cN(0,I)$, $\tilde \bx_j|\bz_j\sim \cN(L\bz_j,\Lambda)$, with $\bz_j$ being a hypothetical latent variable. Then by joint Gaussianity,
\begin{align*}
    \EE[\bz_j|\tilde \bx_j] & = \Sigma_z L^T \Lambda^{-1}\tilde\bx_j, \\
    \EE[\bz_j \bz_j^{\top}|\tilde \bx_j] & = \Sigma_z + \Sigma_zL_k^{\top}\Lambda_k^{-1}\tilde \bx_j\tilde \bx_j^{\top}\Lambda_k^{-1}L_k\Sigma_z, \\
    \EE[\tilde \bx_j\bz_j^{\top} ] & = \tilde \bx_j\tilde \bx_j^{\top} \Lambda_k^{-1}L_k\Sigma_z.
\end{align*}
Plugging these into Eqs.~\eqref{eq:<x>} and~\eqref{eq:<xx>} yields Eqs.~\eqref{eq:<z>},~\eqref{eq:<zz>}, and~\eqref{eq:<xz>}.

\subsubsection{M-step}

Given $X_{\cO}$, the negative posterior expected log-likelihood is
\begin{align*}
    \tr &\cbr{\Sigma^{-1}\conde{XX^\top}} -2\tr\cbr{\Sigma^{-1}\conde{X}\Phi\Theta} + \tr \cbr{\Sigma^{-1}\Theta^\top\Phi^\top\Phi\Theta} \\
    & - 2\tr\cbr{\Sigma^{-1}(\conde{X}-\Theta^\top\Phi^\top)BA^\top} + \tr\cbr{\Sigma^{-1}AB^\top BA^\top } + \lambda \rbr{\tr \cbr{AA^\top} + \tr \cbr{BB^\top}},
\end{align*}
where $\conde{XX^\top} = \sum_{j=1}^n \conde{\bx_j\bx_j^\top}$.
Taking derivatives with respect to $B$ and setting them to 0, we obtain the following update at the $(k+1)$-th iteration:
\[
B_{k+1} = (\conde{X} - \Phi\Theta_k) \Sigma_k^{-1} A_k \rbr{A_k^{\top}\Sigma_k^{-1}A_k + \lambda I_r}^{-1}.
\]
Similarly, we obtain the updates for $A_{k+1}$ and $\Theta_{k+1}$.

We now focus on the update for $L$ and $\Lambda$ now. The complete-data negative log-likelihood is
\begin{align*}
    -\ell_{\text {complete }}(L, \Lambda|X,\{\bz_j\}_{j=1}^n) &= \frac{n}{2} \log |\Lambda|+\frac{1}{2} \sum_{j=1}^n\left\|\bx_j-\bmu_j-L \bz_j\right\|_{\Lambda^{-1}}^2+\frac{1}{2} \sum_{j=1}^n \bz_j^\top \bz_j \\
    &= \frac{n}{2} \log |\Lambda| + \frac 1 2 \fnorm{\Lambda^{-1/2}(X-\Theta_k^\top\Phi^\top - LZ)}^2 + \frac 1 2 \fnorm{Z}^2.
\end{align*}
The posterior expectation is obtained by substituting the sufficient statistics computed in the previous section, yielding
\[
\frac{n}{2} \log |\Lambda| + \frac 1 2  \tr\cbr{\Lambda^{-1}\rbr{\conde{\tilde X\tilde X^\top}- 2 \conde{\tilde X Z^\top}L^\top+ L\conde{ZZ^\top}L^\top}} + \frac 1 2 \tr\cbr{ZZ^\top},
\]
where $\tilde X = X-\Theta_k^\top\Phi^\top$.
The derivatives with respect to $L$ and $\Lambda^{-1}$ are given by
\[
\Lambda^{-1}\conde{ZZ^\top} L - \conde{\tilde XZ^\top} - \conde{XZ^\top},
\]
and
\[
\frac n 2 \Lambda - \frac{1}{2} \rbr{\conde{XX^\top} - 2\conde{XZ^\top}L^\top + L\conde{ZZ^\top} L^\top}.
\]
Setting them to 0 yields the updates for $L_{k+1}$ and $\Lambda_{k+1}$.

\subsection{Step 2}
\label{app:2 deriv}
We rewrite the loss function in Step~2 by reparameterizing $\tilde B \coloneqq B - \Phi\Theta$ as follows.
\[
\begin{aligned}
    \min_{A,\tilde B,\Theta} \|\cP_{\Omega_1}(X^{(1)}-A\tilde B^{\top}-A\Theta^T\Phi^T)\|_F^2
    + \lambda_1\|A\|_F^2 + \lambda_2\|\tilde B_{1:p}\|_F^2 + \alpha\sum_{j=1}^{n-p}\|\tilde \bb_{j+p}-\sum_{l=1}^p\Gamma_l \tilde \bb_{j+p-l})\|^2.
\end{aligned}
\]
Fill in the missing entries in $X$ as follows 
\[
X^*_\Theta=\cP_{\Omega_1}(X^{(1)}-A_k \tilde B_k^{\top}) + \cP_{\Omega_1}^\perp(A_k\Theta_k^{\top}\Phi^{\top}) = \cP_{\Omega_1}(X^{(1)}-A_k B_k^{\top}) + A_k\Theta_k^{\top}\Phi^{\top}.
\]
The loss function for the $\Theta$ update is
\[
\min_Z \tilde F_\Theta(Z|A_k,\tilde B_k,\Theta_k) = \min_{Z} \|X^*_\Theta-A_k^\top Z^\top\Phi^\top\|_F^2.
\]
The partial derivative of $\tilde F_\Theta(\Theta|A_k,\tilde B_k,\Theta_k)$ w.r.t. $\Theta$ is given by
\[
-2 \Phi^\top{X^*}^\top A_k + 2\Phi^\top\Phi\Theta A_k^\top A_k.
\]
Setting it equal to 0 yields the update for $\Theta$.

Consider the update for $\tilde B$ when $A$ and $\Theta$ fixed. 
Let
\[
X^*_B = \cP_{\Omega_1}(X^{(1)}) + \cP_{\Omega_1}^\perp(A_k B_k^{\top}) - A_k \Theta_{k+1}^{\top} \Phi^{\top}
\]
denote the current estimate of the detrended data. The update is obtained by minimizing
\[
\tilde F_B(Z|A_k,\tilde B_k,\Theta_{k+1}) = \fnorm{X^*_B - A_k Z^\top}^2 +\lambda_2\fnorm{Z_{1:p}}^2 + \alpha \sum_{j=1}^{n-p}\left\| \bz_{j+p}-\sum_{l=1}^p\Gamma_l \bz_{j+p-l}\right\|^2. 
\]
where $Z_{1:p} = (\bz_1,\ldots,\bz_p)^\top$. Define
\[
\begin{aligned}
    G(Z)&=\rbr{\sum_{j=0}^p \Gamma_j Z^\top P_n(n-p+j)}\begin{pmatrix}
        I_{n-p} \\ 0
    \end{pmatrix}.
\end{aligned}
\]
Then
\[
\tilde F_B(Z|A_k,\tilde B_k,\Theta_{k+1}) = \fnorm{X^*_B - A_k Z^\top}^2 +\lambda_2\fnorm{Z_{1:p}}^2 + \alpha \left\|G(Z)\right\|^2
\] 
where $\Gamma_0=-I_r$.

Moreover, vectorization yields
\[
\vvec(G(Z)) = C_0 \vvec(Z^\top),
\]
where
\[
C_0 =
\left(\begin{pmatrix} I_{n-p} & 0 \end{pmatrix} \otimes I_r \right)
\left(\sum_{j=0}^p P_n(p-j)\otimes \Gamma_j\right).
\]
Here, $P_n(\cdot)$ denotes the \textit{permutation matrix} which is defined in Table~\ref{tab:app notations}. Note that, intuitively, for any matrix $Y\in\RR^{n\times n'}$, $P_n(k)Y$ rotates the rows of $Y$ upwards by $k$ rows. 

Taking derivatives with respect to $\vvec(Z^\top)$ gives
\[
-[I_n \otimes A_k^\top]\vvec(X^*_B)
+ [I_n \otimes (A_k^\top A_k)] \vvec(Z^\top)
+ \left(\lambda_2 (Q_{1:p}^n \otimes I_r) + \alpha C_0^\top C_0\right)\vvec(Z^\top),
\]
where $Q_{1:p}^n = \operatorname{diag}(\mathbbm{1}(j \le p))$.

Setting the derivative equal to zero yields
\begin{equation}
\begin{aligned}
\vvec(B_{k+1}^{\top})
={}& \Big[J_{A_k}
- \alpha J_{A_k} C_0^\top (I_{(n-p)r} + \alpha C_0 J_{A_k} C_0^\top)^{-1} C_0 J_{A_k}\Big]
[I_n \otimes A_k^\top]\vvec(X^*_B) \\
&+ \vvec(\Theta_{k+1}^\top \Phi^\top),
\end{aligned}
\label{eq:2 B update}
\end{equation}
where
\[
J_A = \left(I_n \otimes (A^\top A) + \lambda_2 (Q_{1:p}^n \otimes I_r)\right)^{-1},
\quad
B_{k+1} = \tilde B_{k+1} + \Phi \Theta_{k+1}.
\]

Note that when $r$ is large, matrix $MJ_{A_k}M^{\top}$ in Eq.~\eqref{eq:2 B update} is still of high dimension, resulting in considerable computational cost. Therefore, we update each $\bb_t$ sequentially for $t=1,\ldots,n$, solving a series of $r$-dimensional linear systems
\begin{align}
    \rbr{A_k^\top A_k + \alpha\sum_{j=\max\cbr{0,p-t+1}}^{\min\cbr{p,n-t}} \Gamma_j^2 + \lambda_2\mathbbm{1}(1\leq t\leq p) I_r }\tilde\bb_t &+ 
    \alpha \rbr{\sum_{j=\max\cbr{0,p-t+1}}^{\min\cbr{p,n-t}} \sum_{\substack{k \neq j, \\ 0 \leq k \leq p}} \Gamma_j\Gamma_k\tilde\bb_{t+j-k}} \nonumber\\
    &= A_k^\top {\bx_B^*}_{,t},
    \label{eq:2 b update}
\end{align}
where ${\bx_B^*}_{,t}$ is the $t$-th column of $X_B^*$. Eq.~\eqref{eq:2 b update} is the direct result from calculating the derivative of $\tilde F_B(Z|A_k,\tilde B_k,\Theta_{k+1})$ w.r.t. $\bz_t$, $t=1,\ldots,n$.

We now consider the update for $A$ obtained by minimizing
\[
\begin{aligned}
    \tilde F_A(Z|A_k,\tilde B_{k+1},\Theta_{k+1}) &= \norm{\cP_{\Omega_1}(X^{(1)}) + \cP_{\Omega_1}^\perp(A_kB_{k+1}^\top) - ZB_{k+1}^\top}_2^2 + \lambda_1\fnorm{Z}^2 \\
    &= \norm{X^*_A - ZB_{k+1}^\top}_2^2 + \lambda_1\fnorm{Z}^2,
\end{aligned}
\]
where $X^*_A=\cP_{\Omega_1}(X^{(1)}) + \cP_{\Omega_1}^\perp(A_kB_{k+1}^\top)$. Taking the derivative and setting it to 0 yields the update for $A$.

\section{On the estimation of $\cbr{\Gamma_j}$}
\label{app:Gamma}
Given $A$ and $B$, minimizing the auto-regression squared error decouples into $p\times r$ linear equations. Let $\Gamma_l=\diag(\bgamma_l)$ where $\bgamma_l$ is a vector of length $r$, and $\tilde\bb_j=\bb_j-\Theta^\top\bphi_j\text{ for }j\in[n]$. Then
\begin{equation}
    \begin{aligned}
        0 &= \diag\rbr{
        \sum_{i=1}^p \Gamma_i \sum_{j=1}^{n-p}\tilde\bb_{j+p-i}\tilde\bb_{j+p-l}^{\top} - \sum_{j=1}^{n-p}\tilde\bb_{j+p}\tilde\bb_{j+p-l}^{\top}
        } \\
        &= \rbr{
        \sum_{i=1}^p \bgamma_i \odot \sum_{j=1}^{n-p}\diag\rbr{\tilde\bb_{j+p-i}\tilde\bb_{j+p-l}^{\top}} - \sum_{j=1}^{n-p}\diag\rbr{\tilde\bb_{j+p}\tilde\bb_{j+p-l}^{\top}}
        },\ l=1,\cdots,p,
    \end{aligned}
    \label{eq:Gamma linear eq}
\end{equation}
where $\odot$ denotes entry-wise multiplication. These linear equations are non-singular if $p\leq\floor{n/2}$ and $\tilde B$ is full-rank. Since the number of variables to be solved for is $p\times r$, \eqref{eq:Gamma linear eq} has a unique solution for $\cbr{\Gamma_l,\ l=1,\cdots,p}$, if exist. 
In practice, we only estimate $cbr{\Gamma_l,\ l=1,\cdots,p}$ for once before Step~2, since iteratively updating them does not significantly increase imputation accuracy and may even introduce unreliable artifacts.

\section{Proof of the theoretical properties}
\label{app:2 proof}
The Step~2 update for $B$ has two variants depending on the magnitude of $r$. Without loss of generality, we restrict attention to the variant with the update specified in Eq.~\ref{eq:2 B update}.

Before proving our results, we present two lemmas from \citet{hastie_matrix_2015}, which will be used subsequently.
\begin{lemma}[\cite{hastie_matrix_2015}]
    \label{lem:quad}
    \[
    H(\bbeta)=\frac1 2\|\by - M\bbeta\|_2^2 + \frac\lambda 2 \|\bbeta\|_2^2,\quad
    \bbeta^* = \argmin_{\bbeta} H(\bbeta).
    \]
    Then
    \[
    H(\bbeta) -  H(\bbeta^*) = \frac1 2(\bbeta-\bbeta^*)^{\top}(M^{\top}M+\lambda I)(\bbeta-\bbeta^*) = \frac1 2\|M(\bbeta-\bbeta^*)\|_2^2 + \frac\lambda 2\|\bbeta-\bbeta^*\|_2^2.
    \]
\end{lemma}

\begin{lemma}[\citet{hastie_matrix_2015}]
    \label{lem:surrogate}
    $\forall X,\bar Z, Z\in\RR^{m\times n}$, $\|\cP_\Omega(X-Z)\|_F^2\leq \|\cP_\Omega(X)+\cP_\Omega^\perp(\bar Z)-Z\|_F^2$.
\end{lemma}
\begin{proof}
    $\forall (i,j)\in\cbr{(i,j):i\in[m],j\in[n]}$, if $(i,j)\in\Omega$, then $\cP_\Omega(X-Z)_{ij}=X_{ij}-Z_{ij}=\cP_\Omega(X)_{ij}-Z_{ij}$; if $(i,j)\notin\Omega$, then $\cP_\Omega(X-Z)_{ij}=0$. Therefore, 
    \[
    \sum_{(i,j)}\rbr{\cP_\Omega(X-Z)_{ij}}^2 \leq \sum_{(i,j)}\rbr{\cP_\Omega(X)_{ij}+\cP_\Omega^\perp(\bar Z)_{ij}-Z_{ij}}^2. 
    \]
    The equality holds when $\bar Z=Z$. \qedhere
\end{proof}

We rewrite the loss function in the following form by reparameterizing $\tilde B \coloneq B - \Phi\Theta$.
\[
\begin{aligned}
    \min_{A,\tilde B,\Theta} \|\cP_{\Omega_1}(X^{(1)}-A\tilde B^{\top}-A\Theta^T\Phi^T)\|_F^2
    + \lambda_1\|A\|_F^2 + \lambda_2\|\tilde B_{1:p}\|_F^2 + \alpha\sum_{j=1}^{n-p}\|\tilde \bb_{j+p}-\sum_{l=1}^p\Gamma_l \tilde \bb_{j+p-l})\|^2.
\end{aligned}
\]
In this section, we define loss function
\begin{align*}
    \bar F_2(A,\tilde B,\Theta) & = \|\cP_{\Omega_1}(X^{(1)}-A\tilde B^{\top}-A\Theta^T\Phi^T)\|_F^2
    + \lambda_1\|A\|_F^2 + \lambda_2\|\tilde B_{1:p}\|_F^2 + \\
    & \alpha\sum_{j=1}^{n-p}\|\tilde\bb_{j+p}-\sum_{l=1}^p\Gamma_l \tilde\bb_{j+p-l})\|^2,
\end{align*}
such that $\bar F_2(A,\tilde B,\Theta) = \bar F_2(A,B-\Phi\Theta,\Theta) = F_2(A,B,\Theta)$, for any $(A,B,\Theta)$.

\subsection{Proof of Theorem~3.1}
\label{app:2 decrease}
\begin{proof}
Define majorization functions of $\bar F_2(A,\tilde B,\Theta)$:
\[
\begin{aligned}
    \tilde F_A(Z|A,\tilde B,\Theta) & =\fnorm{\cP_{\Omega_1}(X^{(1)}-A\Theta^\top\Phi^\top) + \cP_{\Omega_1}^\perp(A\tilde B^\top)-Z\tilde B^\top}^2 + \lambda_1\fnorm{Z}^2 + C_A(\tilde B,\Theta), \\
    \tilde F_B(Z|A,\tilde B,\Theta) & = \fnorm{\cP_{\Omega_1}(X^{(1)}-A\Theta^\top\Phi^\top) + \cP_{\Omega_1}^\perp(A\tilde B^\top)-AZ^\top}^2 + \lambda_2\fnorm{Z_{1:p}}^2 + \\ 
    & \alpha \sum_{j=1}^{n-p} \norm{\bz_{j+p} - \sum_{k=1}^p \Gamma_k\bz_{j+p-k} }^2 + C_B(A,\Theta), \\
    \tilde F_\Theta(Z|A,\tilde B,\Theta) & = \fnorm{\cP_{\Omega_1}(X^{(1)} - A\tilde B^\top ) + \cP_{\Omega_1}^\perp(A\Theta^\top\Phi^\top) - AZ^\top\Phi^\top}^2 + C_\Theta(\tilde B).
\end{aligned}
\]
Observe that
\[
\tilde F_A(Z|A,\tilde B,\Theta) \geq \bar F_2(Z,\tilde B,\Theta).
\]
By Lemma~\ref{lem:surrogate}, the equality holds when $Z=A$. Similarly,
\[
\tilde F_B(Z|A,\tilde B,\Theta) \geq \bar F_2(A,Z,\Theta), 
\text{ and }
\tilde F_\Theta(Z|A,\tilde B,\Theta) \geq \bar F_2(A,\tilde B,Z),
\]
where the equalities hold when $Z=\tilde B$, and $Z=\Theta$, respectively. Note that 

Since $\Theta_{k+1} = \argmin_{Z} \tilde F_\Theta(Z|A_k,B_k,\Theta_k)$, 
\[
\bar F_2(\Theta_{k+1}|A_k,\tilde B_k,\Theta_k) \leq \tilde F_\Theta(\Theta_{k+1}|A_k,\tilde B_k,\Theta_k) \leq \tilde F_\Theta(\Theta_k|A_k,\tilde B_k,\Theta_k) = \bar F_2(\Theta_k|A_k,\tilde B_k,\Theta_k). 
\]
Similarly, we can prove $\bar F_2(A_k,\tilde B_k,\Theta_{k+1}) \geq \bar F_2(A_k,\tilde B_{k+1},\Theta_{k+1}) \geq \bar F_2(A_{k+1},\tilde B_{k+1},\Theta_{k+1})$. Therefore, $F_2(A_k,B_k,\Theta_{k}) \geq F_2(A_{k+1},B_{k+1},\Theta_{k+1})$.
\end{proof}

\subsection{Proof of Lemma~3.2}
\begin{proof}
To prove this Lemma, we first prove the following Lemma~\ref{lem:2 low} which quantifies the lower bound of $\Delta_k$.
\begin{lemma}[Lower bound of $\Delta_k$]
    \label{lem:2 low}
    \begin{align*}
        \Delta_k \geq & \fnorm{A(\Theta_k-\Theta_{k+1})^\top\Phi^\top}^2 \\
        & + \fnorm{A_k(\tilde B_k-\tilde B_{k+1})^\top}^2 + \lambda_2 \fnorm{(\tilde B_k-\tilde B_{k+1})_{1:p}}^2 + \alpha \fnorm{G(\tilde B_k-\tilde B_{k+1})}^2 \\
        & + \fnorm{(A_k-A_{k+1})\tilde B_{k+1}^\top}^2 + \lambda_1 \fnorm{(A_k-A_{k+1})}^2. 
    \end{align*}
    where for any $Z\in\RR^{n\times r}$, $G(Z)=\rbr{\sum_{j=0}^p \Gamma_j Z^{\top} P_n(n-p+j)}\begin{pmatrix}
            I_{n-p} & 0
        \end{pmatrix}^\top$.
\end{lemma}
\begin{proof}[Proof of Lemma~\ref{lem:2 low}]
By Lemma~\ref{lem:quad},
\[
\begin{aligned}
    \bar F_2(A_k,\tilde B_k,\Theta_k) &- \bar F_2(A_k,\tilde B_k,\Theta_{k+1}) = \tilde F_\Theta(\Theta_k|A_k,\tilde B_k,\Theta_k) - \tilde F_\Theta(\Theta_{k+1}|A_k,\tilde B_k,\Theta_k) \\
    & \overset{\mathrm{Lemma~\ref{lem:quad}}}{=\mathrel{\mkern-3mu}=} \fnorm{A(\Theta_k-\Theta_{k+1})^\top\Phi^\top}^2.
\end{aligned}
\]
Similarly,
\[
\begin{aligned}
    \bar F_2(A_k, & \tilde B_k,\Theta_{k+1}) - \bar F_2(A_k,\tilde B_{k+1},\Theta_{k+1}) \\
    \geq &  \tilde F_B(\tilde B_k|A_k,\tilde B_k,\Theta_{k+1}) - \tilde F_B(\tilde B_{k+1}|A_k,\tilde B_k,\Theta_{k+1})\\
    \overset{\mathrm{Lemma~\ref{lem:quad}}}{=\mathrel{\mkern-3mu}=} & \vvec^\top(\tilde B_k^\top-\tilde B_{k+1}^\top) \rbr{I_n\otimes A_k^\top A_k + \lambda_2 Q^n_{1:p} \otimes I_r + \alpha C_0^\top C_0} \vvec(\tilde B_k^\top-\tilde B_{k+1}^\top) \\
     = & \fnorm{A_k(\tilde B_k-\tilde B_{k+1})^\top}^2 + \lambda_2 \fnorm{(\tilde B_k-\tilde B_{k+1})_{1:p}}^2 + \alpha \fnorm{G(\tilde B_k-\tilde B_{k+1})}^2,
\end{aligned}
\]

\[
\begin{aligned}
    \bar F_2(A_k, & \tilde B_{k+1},\Theta_{k+1}) -  \bar F_2(A_{k+1},\tilde B_{k+1},\Theta_{k+1}) \\
    & \geq  \tilde F_A(A_k|A_k,B_{k+1},\Theta_{k+1}) - \tilde F_A(A_{k+1}|A_k,\tilde B_{k+1},\Theta_{k+1})\\
    &\overset{\mathrm{Lemma~\ref{lem:quad}}}{=\mathrel{\mkern-3mu}=} \fnorm{(A_k-A_{k+1})\tilde B_{k+1}^\top}^2 + \lambda_1 \fnorm{(A_k-A_{k+1})}^2.
\end{aligned}
\]
Therefore,
\begin{align*}
    \Delta_k \geq & \fnorm{A(\Theta_k-\Theta_{k+1})^\top\Phi^\top}^2 \\
    & + \fnorm{A_k(\tilde B_k-\tilde B_{k+1})^\top}^2 + \lambda_2 \fnorm{(\tilde B_k-\tilde B_{k+1})_{1:p}}^2 + \alpha \fnorm{G(\tilde B_k-\tilde B_{k+1})}^2 \\
    & + \fnorm{(A_k-A_{k+1})\tilde B_{k+1}^\top}^2 + \lambda_1 \fnorm{(A_k-A_{k+1})}^2. 
\end{align*}
\end{proof}

Lemma~\ref{lem:2 low} shows that $\Delta_k$ has a lower bound. This guarantees that $\{F_2(A_k,B_k,\Theta_k)\}_{k\geq1}$ is a strictly decreasing sequence unless the they reach a fixed point. 
Let $f$ denote the update function of Algorithm~2, so that$(A_{k+1},B_{k+1},\Theta_{k+1}) = f(A_k,B_k,\Theta_k),\ k=1,\ldots,K$.
If $\Delta_k = 0$, that is, $F_2(A_{k+1},B_{k+1},\Theta_{k+1}) = F_2(A_k,B_k,\Theta_k)$, then by the proof of Lemma~\ref{lem:2 low}, it follows that
\begin{align*}
    &\tilde F_\Theta(\Theta_k|A_k,B_k,\Theta_k) = \tilde F_\Theta(\Theta_{k+1}|A_k,B_k,\Theta_k), \\
    &\tilde F_B(B_k|A_k,B_k,\Theta_{k+1}) = \tilde F_B(B_{k+1}|A_k,B_k,\Theta_{k+1}), \\
    &\tilde F_A(A_k|A_k,B_{k+1},\Theta_{k+1}) = \tilde F_A(A_{k+1}|A_k,B_{k+1},\Theta_{k+1}).
\end{align*}
Because each block update for \(A\), \(B\), and \(\Theta\) is a strictly convex optimization problem, each update has a unique minimizer. Hence,
\[
(A_k, B_k, \Theta_k) = (A_{k+1}, B_{k+1}, \Theta_{k+1}).
\]
Conversely, suppose that $(A_k,B_k,\Theta_k)$ is a fixed point. Then 
\begin{align*}
    &\tilde F_\Theta(\Theta_k|A_k,B_k,\Theta_k) = \tilde F_\Theta(\Theta_{k+1}|A_k,B_k,\Theta_k), \\
    &\tilde F_B(B_k|A_k,B_k,\Theta_{k+1}) = \tilde F_B(B_{k+1}|A_k,B_k,\Theta_{k+1}), \\
    &\tilde F_A(A_k|A_k,B_{k+1},\Theta_{k+1}) = \tilde F_A(A_{k+1}|A_k,B_{k+1},\Theta_{k+1}).
\end{align*}
Therefore,  by convexity, $(A_k,B_k,\Theta_k) = (A_{k+1},B_{k+1},\Theta_{k+1})$, which implies $\Delta_k=0$.
\end{proof}

\subsection{Proof of Theorem~3.3}
\label{app:2 up}
\begin{proof}
Since $\Delta_k\geq 0$ and $\sum_{k=1}^\infty \Delta_k = F_2(A_1,B_1,\Theta_1) - F_2^\infty = \bar F_2(A_1,\tilde B_1,\Theta_1) - \bar F_2^\infty < \infty$, it follows that $\lim_{k\rightarrow \infty}\Delta_k = 0$.

Observe that
\[
\sum_{k=1}^K \rbr{F_2(A_k, B_k, \Theta_k) - F_2(A_{k+1}, B_{k+1}, \Theta_{k+1})} = \sum_{k=1}^K \Delta_k \geq K(\min_{1\leq k\leq K} \Delta_k),
\]
where the LHS
\begin{align*}
F_2(A_1,B_1,\Theta_1) - F_2(A_{K+1},B_{K+1},\Theta_{K+1}) \leq F_2(A_1,B_1,\Theta_1) - F_2^\infty.
\end{align*}
Therefore
\[
\min_{1\leq k\leq K} \Delta_k \leq \frac{F_2(A_1,B_1,\Theta_1) - F_2^\infty}{K}. \qedhere
\]
\end{proof}

\subsection{Proof of Theorem~3.4}
\label{app:2 stationary point}
\begin{proof}
We first prove the equivalence between fixed point and stationary point.
\begin{proposition}
    \label{prop:stationary and fixed point}
    An equivalent restatement of the stationary point of the Step~2 loss function is 
    \[
    A_* = \argmin_Z \tilde F_A(Z|A_*,\tilde B_*,\Theta_*), \ 
    B_* = \argmin_Z\tilde F_B(Z|A_*,\tilde B_*,\Theta_*), \ 
    \Theta_* = \argmin_Z\tilde F_\Theta(Z|A_*,\tilde B_*,\Theta_*),
    \]
    i.e. $(A_*,\tilde B_*,\Theta_*)$ is a fixed point.
\end{proposition}
\begin{proof}[Proof of Proposition~\ref{prop:stationary and fixed point}]
    Without loss of generality, we restrain our attention to the proof for $A_*$. The result extends readily to $\tilde B_*$ and $\Theta_*$. If $(A_*, B_*,\Theta_*)$ is a stationary point of $F_2$, then $(A_*,\tilde B_*,\Theta_*)$ is a stationary point of $\bar F_2$. Hence, 
    $$\nabla\tilde F_A(A_*|A_*,\tilde B_*,\Theta_*)=\nabla_A \bar F_2(A_*,\tilde B_*,\Theta_*)=0.$$ 
    It follows that $A_*$ minimizes $\tilde F_A(\cdot|A_*,\tilde B_*,\Theta_*)$, i.e. $A_* = \argmin\tilde F_A(Z|A_*,\tilde B_*,\Theta_*)$.
    
    Conversely, by the definitions of $\tilde F_A(Z|A_*,\tilde B_*,\Theta_*)$ and $\bar F_2$, $\tilde F_A(Z|A_*,\tilde B_*,\Theta_*)$ is tangent to $\bar F_2(Z,\tilde B_*,\Theta_*)$ as a function of $Z$ at $A_*$. Therefore, 
    $$\nabla_A \bar F_2(A_*, B_*,\Theta_*) = \nabla_A \bar F_2(A_*,\tilde B_*,\Theta_*)=\nabla\tilde F_A(A_*|A_*,\tilde B_*,\Theta_*)=0. \qedhere$$
\end{proof}

Since $\Delta_k\geq 0$ and $\sum_{k=1}^\infty \Delta_k = F_2(A_1,B_1,\Theta_1) - F_2^\infty < \infty$, it follows that
\[
\Delta_k = F_2(A_k,B_k,\Theta_k) - F_2(A_{k+1},B_{k+1},\Theta_{k+1}) \rightarrow 0.
\]
Suppose $(A_*,\tilde B_*,\Theta_*)$ is a limit point of $\cbr{(A_k,\tilde B_k,\Theta_k)}$. Then $\bar F_2(A_k,\tilde B_k,\Theta_k) \rightarrow \bar F_2(A_*,\tilde B_*,\Theta_*)$. Therefore,
$\Delta_* \coloneqq \bar F_2(A_*,\tilde B_*,\Theta_*) - \bar F_2^\infty = 0$. By Lemma~3.2, $(A_*,\tilde B_*,\Theta_*)$ is a fixed point of the algorithm, implying that $(A_*,B_*,\Theta_*)$ is a stationary point of the objective function.
\end{proof}

\section{Illustration of uncertainty quantification method}

Figure~\ref{fig:uq diagram} illustrates the sample-splitting procedure and the corresponding exchangeability properties in proposed uncertainty quantification method.

\begin{figure}[h!]
\centering
\begin{tikzpicture}[
    every node/.style={anchor=center}, 
    nodes={font=\scriptsize},
    label/.style={font=\small, inner sep=1pt, rounded corners},
    >={Stealth[round]}, 
    highlight31/.style={draw=red!40!black, very thick, rounded corners, inner sep=1pt},
    highlight32/.style={draw=red!70!black, very thick, rounded corners, inner sep=1pt},
    highlight33/.style={draw=red, very thick, rounded corners, inner sep=1pt},
    highlight34/.style={draw=red!60!yellow, very thick, rounded corners, inner sep=1pt},
    highlight35/.style={draw=red!40!yellow, very thick, rounded corners, inner sep=1pt},
    highlight21/.style={thick, rounded corners, fill=blue!20, inner sep=2pt},
    highlight22/.style={thick, rounded corners, fill=yellow!20, inner sep=2pt},
    highlight1/.style={
        pattern=north east lines, 
        pattern color=gray!60!black,       
        thick, rounded corners, inner sep=2pt}
]
\matrix (m1) [matrix of nodes, nodes in empty cells,
             nodes={minimum width=.5cm, minimum height=.5cm, text depth=0.1ex, text height=1.2ex, draw},
             column sep=0.1cm, row sep=0.1cm] 
{
   &  &  &  &  &  &  &  \\
   &  &  &  &  &  &  &  \\
   &  &  &  &  &  &  &  \\
   &  &  &  &  &  &  &  \\
   &  &  &  &  &  &  &  \\
};

\begin{pgfonlayer}{background}
  \node[highlight1, fit=(m1-1-1)(m1-1-1)] {};
  \node[highlight1, fit=(m1-5-1)(m1-5-1)] {};
  \node[highlight1, fit=(m1-2-2)(m1-3-2)] {};
  \node[highlight1, fit=(m1-4-3)(m1-4-3)] {};
  \node[highlight1, fit=(m1-1-4)(m1-5-4)] {};
  \node[highlight1, fit=(m1-1-6)(m1-5-6)] {};
  \node[highlight1, fit=(m1-3-7)(m1-3-7)] {};
  \node[highlight1, fit=(m1-5-7)(m1-5-7)] {};
  \node[highlight1, fit=(m1-2-8)(m1-2-8)] {};
\end{pgfonlayer}

\node[above=.1 of m1] {\textbf{(A). Missing entries (test set).}};

\node[left=.3 of m1-3-1, rotate=90, anchor=center] (rlabel) {\textcolor{black}{Wavelength}};

\node[below=.5pt of m1-5-5.south west] (clabel) {Time};


\matrix (m2) [matrix of nodes, nodes in empty cells,
             nodes={minimum width=.5cm, minimum height=.5cm, text depth=0.1ex, text height=1.2ex, draw},
             column sep=0.1cm, row sep=0.1cm, right=1.7cm of m1] 
{
   &  &  &  &  &  &  &  \\
   &  &  &  &  &  &  &  \\
   &  &  &  &  &  &  &  \\
   &  &  &  &  &  &  &  \\
   &  &  &  &  &  &  &  \\
};

\begin{pgfonlayer}{background}
  \node[highlight21, fit=(m2-3-1)(m2-3-1)] {};
  \node[highlight21, fit=(m2-4-2)(m2-4-2)] {};
  \node[highlight22, fit=(m2-1-3)(m2-5-3)] {};
  \node[highlight21, fit=(m2-1-5)(m2-1-5)] {};
  \node[highlight21, fit=(m2-2-7)(m2-2-7)] {};
  \node[highlight21, fit=(m2-5-8)(m2-5-8)] {};
  \node[highlight1, fit=(m2-1-1)(m2-1-1)] {};
  \node[highlight1, fit=(m2-5-1)(m2-5-1)] {};
  \node[highlight1, fit=(m2-2-2)(m2-3-2)] {};
  \node[highlight1, fit=(m2-4-3)(m2-4-3)] {};
  \node[highlight1, fit=(m2-1-4)(m2-5-4)] {};
  \node[highlight1, fit=(m2-1-6)(m2-5-6)] {};
  \node[highlight1, fit=(m2-3-7)(m2-3-7)] {};
  \node[highlight1, fit=(m2-5-7)(m2-5-7)] {};
  \node[highlight1, fit=(m2-2-8)(m2-2-8)] {};
\end{pgfonlayer}

\node[above=.1 of m2] {\textbf{(B). Calibration and training sets.}};

\matrix (m3) [matrix of nodes, nodes in empty cells,
             nodes={minimum width=.5cm, minimum height=.5cm, text depth=0.1ex, text height=1.2ex, draw},
             column sep=0.1cm, row sep=0.1cm, below=1cm of m1] 
{
   &  &  &  &  &  &  &  \\
   &  &  &  &  &  &  &  \\
   &  &  &  &  &  &  &  \\
   &  &  &  &  &  &  &  \\
   &  &  &  &  &  &  &  \\
};

\begin{pgfonlayer}{background}
  \node[highlight21, fit=(m3-3-1)(m3-3-1)] {};
  \node[highlight21, fit=(m3-4-2)(m3-4-2)] {};
  \node[highlight22, fit=(m3-1-3)(m3-5-3)] {};
  \node[highlight21, fit=(m3-1-5)(m3-1-5)] {};
  \node[highlight21, fit=(m3-2-7)(m3-2-7)] {};
  \node[highlight21, fit=(m3-5-8)(m3-5-8)] {};
  \node[highlight1, fit=(m3-1-1)(m3-1-1)] {};
  \node[highlight1, fit=(m3-5-1)(m3-5-1)] {};
  \node[highlight1, fit=(m3-2-2)(m3-3-2)] {};
  \node[highlight1, fit=(m3-4-3)(m3-4-3)] {};
  \node[highlight1, fit=(m3-1-4)(m3-5-4)] {};
  \node[highlight1, fit=(m3-1-6)(m3-5-6)] {};
  \node[highlight1, fit=(m3-3-7)(m3-3-7)] {};
  \node[highlight1, fit=(m3-5-7)(m3-5-7)] {};
  \node[highlight1, fit=(m3-2-8)(m3-2-8)] {};
  \node[highlight31, fit=(m3-1-1)(m3-1-1)] {};
  \node[highlight31, fit=(m3-1-5)(m3-1-5)] {};
  \node[highlight32, fit=(m3-2-2)(m3-2-2)] {};
  \node[highlight32, fit=(m3-2-7)(m3-2-8)] {};
  \node[highlight33, fit=(m3-3-1)(m3-3-2)] {};
  \node[highlight33, fit=(m3-3-7)(m3-3-7)] {};
  \node[highlight34, fit=(m3-4-2)(m3-4-2)] {};
  \node[highlight35, fit=(m3-5-1)(m3-5-1)] {};
  \node[highlight35, fit=(m3-5-7)(m3-5-8)] {};
\end{pgfonlayer}

\node[above=.1 of m3] {\textbf{(C). Scattered missingness exchangeability.}};

\matrix (m4) [matrix of nodes, nodes in empty cells,
             nodes={minimum width=.5cm, minimum height=.5cm, text depth=0.1ex, text height=1.2ex, draw},
             column sep=0.1cm, row sep=0.1cm, below=1cm of m2] 
{
   &  &  &  &  &  &  &  \\
   &  &  &  &  &  &  &  \\
   &  &  &  &  &  &  &  \\
   &  &  &  &  &  &  &  \\
   &  &  &  &  &  &  &  \\
};

\begin{pgfonlayer}{background}
  \node[highlight21, fit=(m4-3-1)(m4-3-1)] {};
  \node[highlight21, fit=(m4-4-2)(m4-4-2)] {};
  \node[highlight22, fit=(m4-1-3)(m4-5-3)] {};
  \node[highlight21, fit=(m4-1-5)(m4-1-5)] {};
  \node[highlight21, fit=(m4-2-7)(m4-2-7)] {};
  \node[highlight21, fit=(m4-5-8)(m4-5-8)] {};
  \node[highlight1, fit=(m4-1-1)(m4-1-1)] {};
  \node[highlight1, fit=(m4-5-1)(m4-5-1)] {};
  \node[highlight1, fit=(m4-2-2)(m4-3-2)] {};
  \node[highlight1, fit=(m4-4-3)(m4-4-3)] {};
  \node[highlight1, fit=(m4-1-4)(m4-5-4)] {};
  \node[highlight1, fit=(m4-1-6)(m4-5-6)] {};
  \node[highlight1, fit=(m4-3-7)(m4-3-7)] {};
  \node[highlight1, fit=(m4-5-7)(m4-5-7)] {};
  \node[highlight1, fit=(m4-2-8)(m4-2-8)] {};
  \node[highlight31, fit=(m4-1-3)(m4-1-4)] {};
  \node[highlight31, fit=(m4-1-6)(m4-1-6)] {};
  \node[highlight32, fit=(m4-2-3)(m4-2-4)] {};
  \node[highlight32, fit=(m4-2-6)(m4-2-6)] {};
  \node[highlight33, fit=(m4-3-3)(m4-3-4)] {};
  \node[highlight33, fit=(m4-3-6)(m4-3-6)] {};
  \node[highlight34, fit=(m4-4-3)(m4-4-4)] {};
  \node[highlight34, fit=(m4-4-6)(m4-4-6)] {};
  \node[highlight35, fit=(m4-5-3)(m4-5-4)] {};
  \node[highlight35, fit=(m4-5-6)(m4-5-6)] {};
\end{pgfonlayer}

\node[above=.1 of m4] {\textbf{(D). Downtime missingness exchangeability.}};
\end{tikzpicture} 
\caption{Illustration of exchangeability for absolute residuals $R_{ij}$. (A) The shaded entries are missing in the input SSI matrix, constituting the test set; (B) The blue and yellow entries are randomly selected to form the calibration sets for scattered and downtime missingness, respectively. The remaining white entries form the training set; (C) Scattered missingness exchangeability: within the same row, the entries in scattered test set and scattered calibration set are exchangeable; (D) Downtime missingness exchangeability: within the same row, the entries in the downtime test set and downtime calibration set are exchangeable.}
\label{fig:uq diagram}
\end{figure} 

\section{Supplementary experimental results}
\label{sec:supplement}

\subsection{Synthetic data analysis}
\begin{figure}[t]
    \centering
    \includegraphics[width=0.9\linewidth]{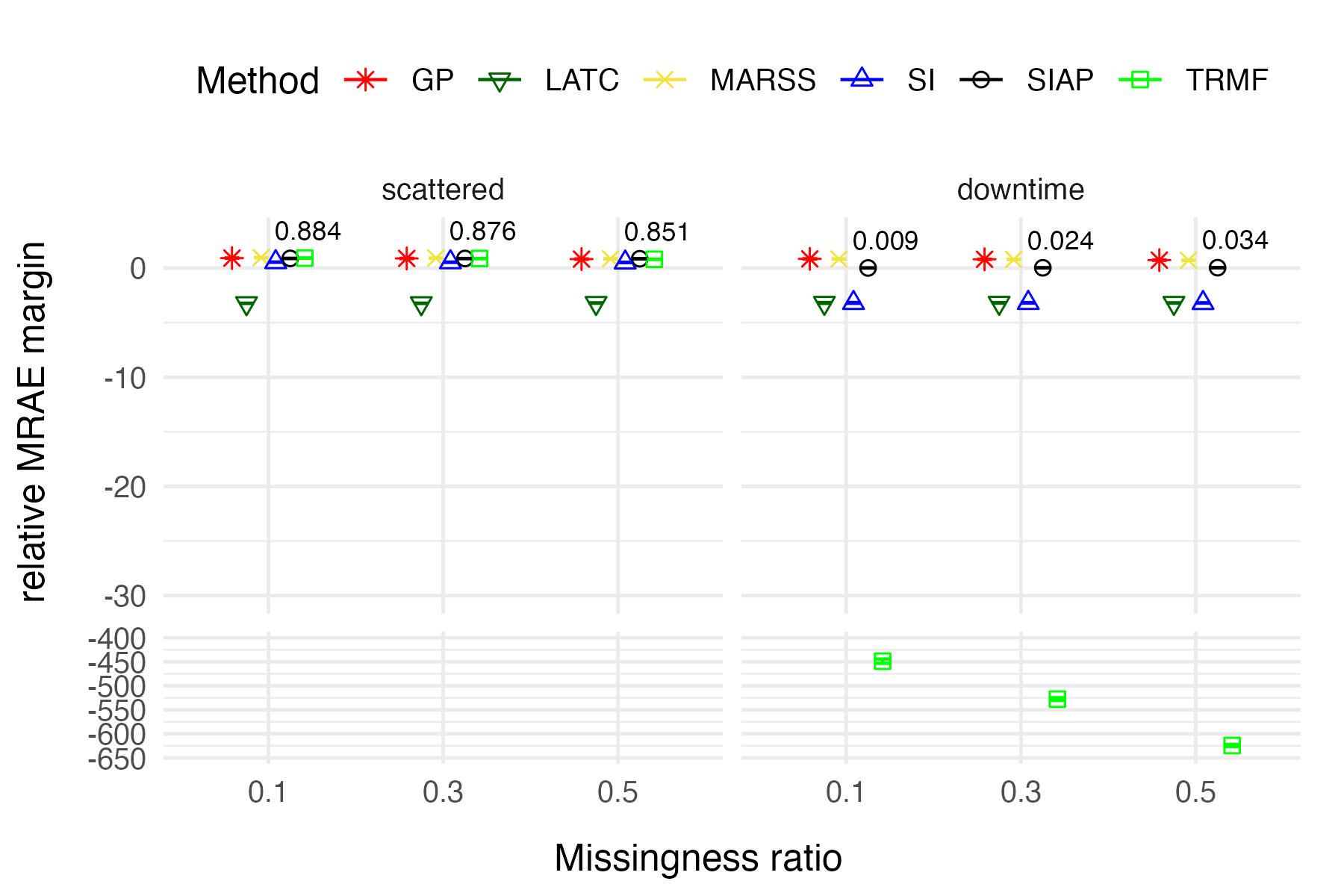}
    \caption{Imputation accuracy of selected methods with varying missingness ratio, and without row-wise first-order difference on GP and MARSS. The scatter points are the mean of the test set relative MRAE margin and the error bars give the 95\% confidence intervals. The MRAE margin is calculated relative to periodic cubic spline fitting.}
    \label{fig:mrae margin nodiff}
\end{figure}

\begin{figure}[t]
    \centering
    \begin{subfigure}[b]{\linewidth}
        \includegraphics[width=\linewidth]{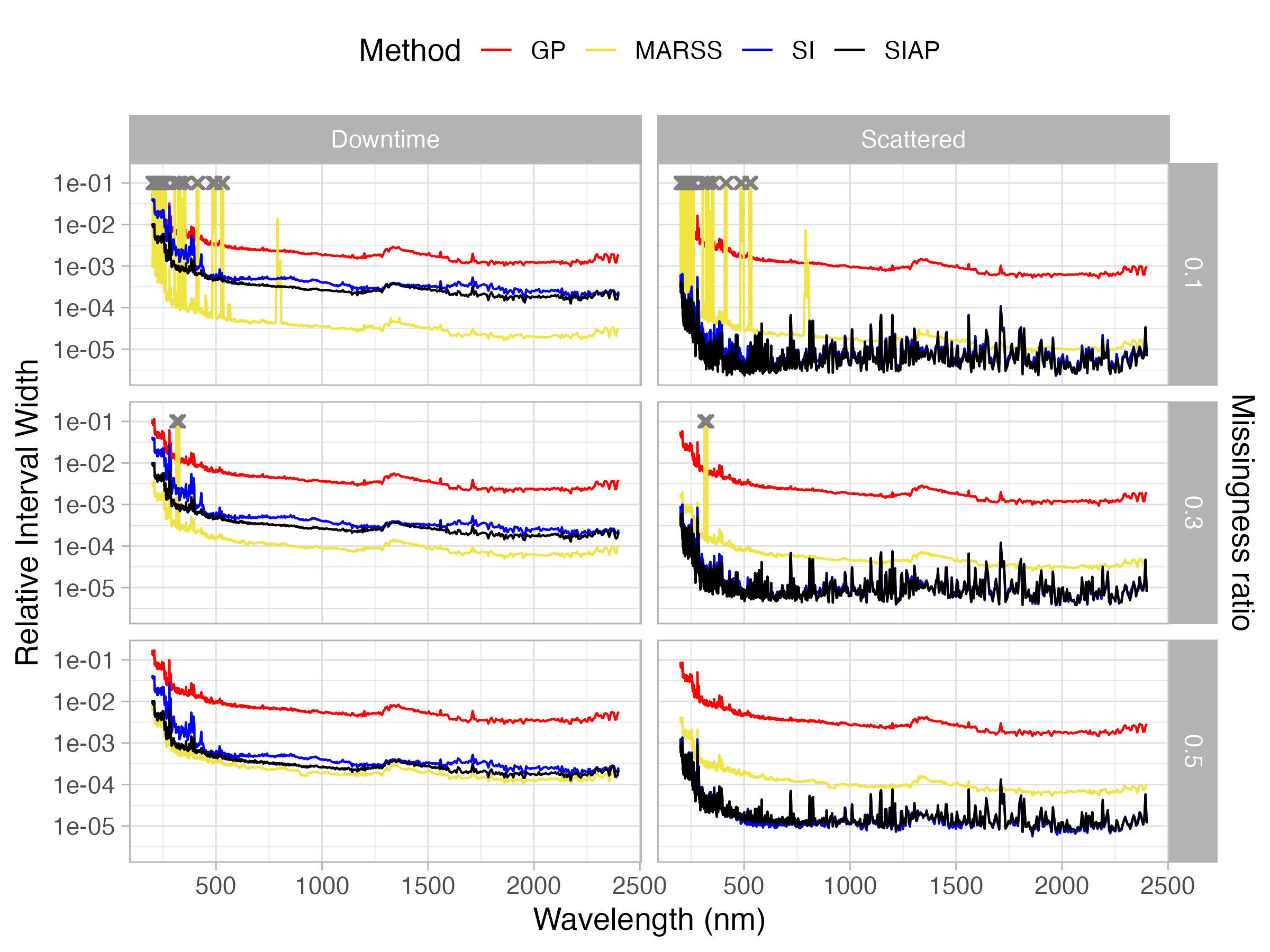}
        \caption{With row-wise difference.}
    \end{subfigure}
    \begin{subfigure}[b]{\linewidth}
        \includegraphics[width=\linewidth]{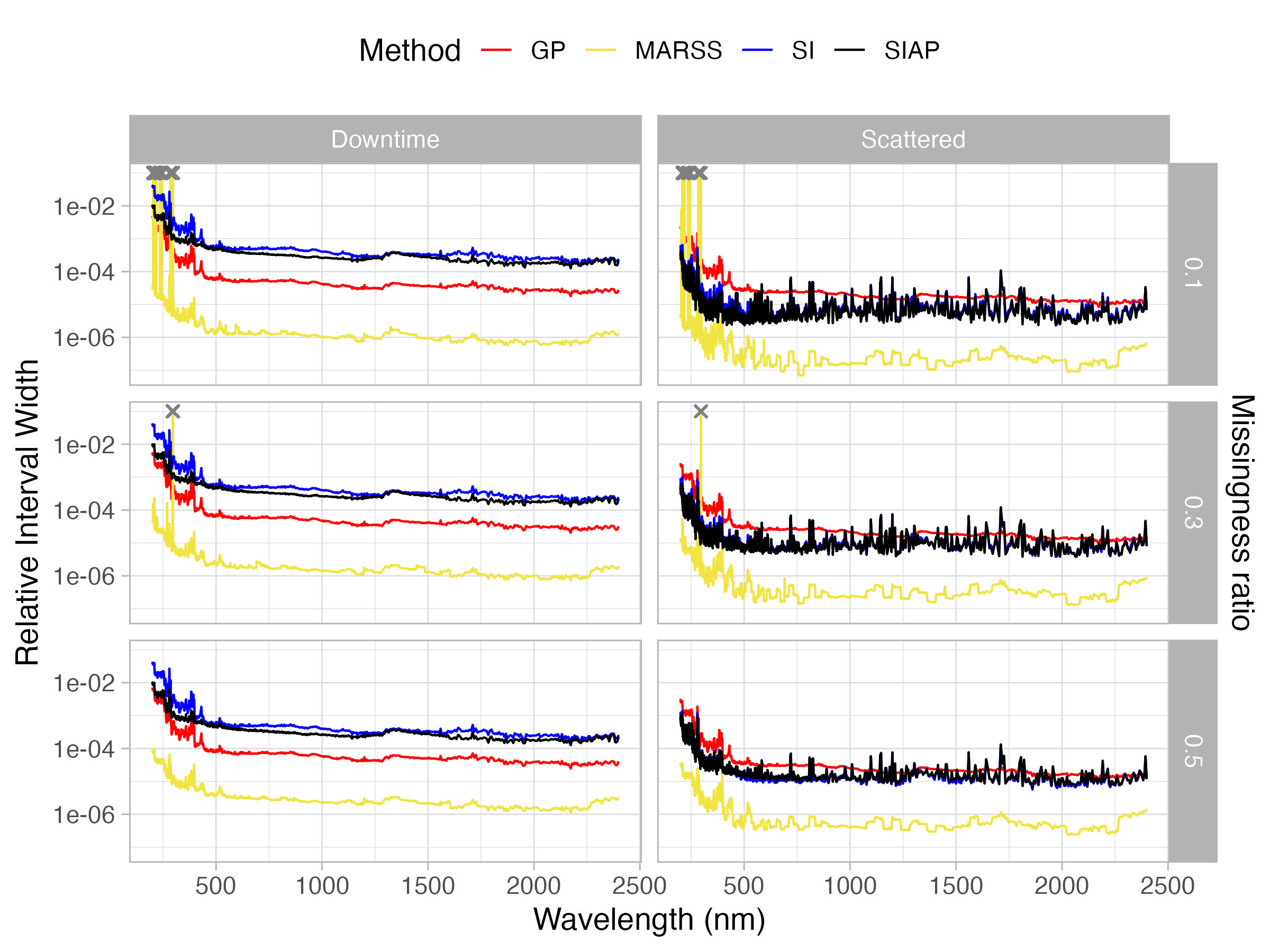}
        \caption{Without row-wise difference.}
    \end{subfigure}
    \caption{Spectral distribution of the relative prediction half-interval widths w.r.t. the global mean values of the synthetic SSI data. There are some large values for MARSS model, which are indicated by the grey crosses. 
    }
    \label{fig:simulation rel interval length}
\end{figure}

Figure~\ref{fig:mrae margin nodiff} compares the imputation accuracy of selected methods without applying row-wise first-order difference on GP and MARSS.
Figure~\ref{fig:simulation rel interval length} displays the spectral distribution of the relative prediction interval lengths.

\subsection{Real data experiment}

\begin{figure}[t]
    \centering
    \includegraphics[width=0.9\linewidth]{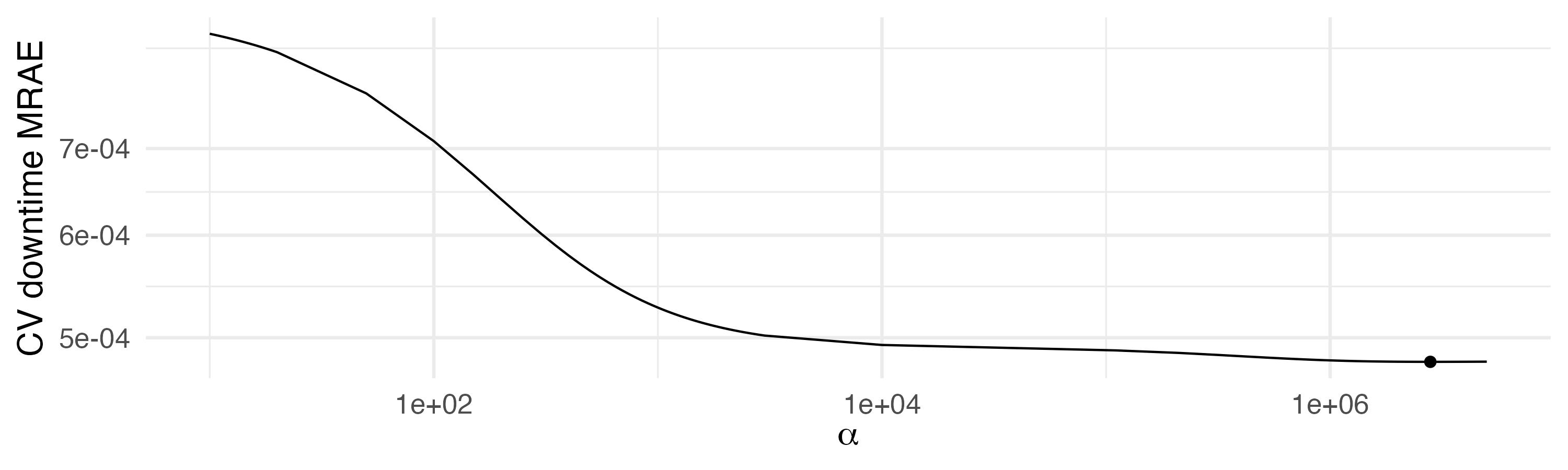}
    \caption{Cross-validation error (downtime MRAE) evaluated over a range of $\alpha$ values, chosen to be on the same scale as the leading eigenvalue of $A^\top A$. The optimal $\alpha$ is marked by the point.}
    \label{fig:alpha path}
\end{figure}

We tune \texttt{SIAP} hyperparameters sequentially. First, \(\lambda\) is selected by 10-fold cross-validation on observed entries, choosing the value minimizing scattered missingness MRAE. The rank is set automatically by \texttt{SoftImpute-ALS} with maximum \(\min\{n,m\}\).
Next, scattered missing entries are imputed with the retrained Step 1 model. Non-missing columns are split into 10 folds, and the model is retrained without AR regularization to select \(\lambda_1, \lambda_2\) minimizing downtime missingness MRAE. Rank is again induced by these parameters.
The order \(p\) is chosen based on the Bayes information criteria (BIC) on the time series \(\{\mathbf{b}_t\}_{t=1}^n\) from the model retrained with optimal \(\lambda_1, \lambda_2\).
Finally, with \(\lambda=11.7\), \(\lambda_1=\lambda_2=8.1\), and \(p=6\), \(\alpha\) is selected by cross-validation downtime MRAE. Figure~\ref{fig:alpha path} shows the MRAE path used to determine the optimal \(\alpha\).

\begin{figure}[t]
    \centering
    \includegraphics[width=0.8\linewidth]{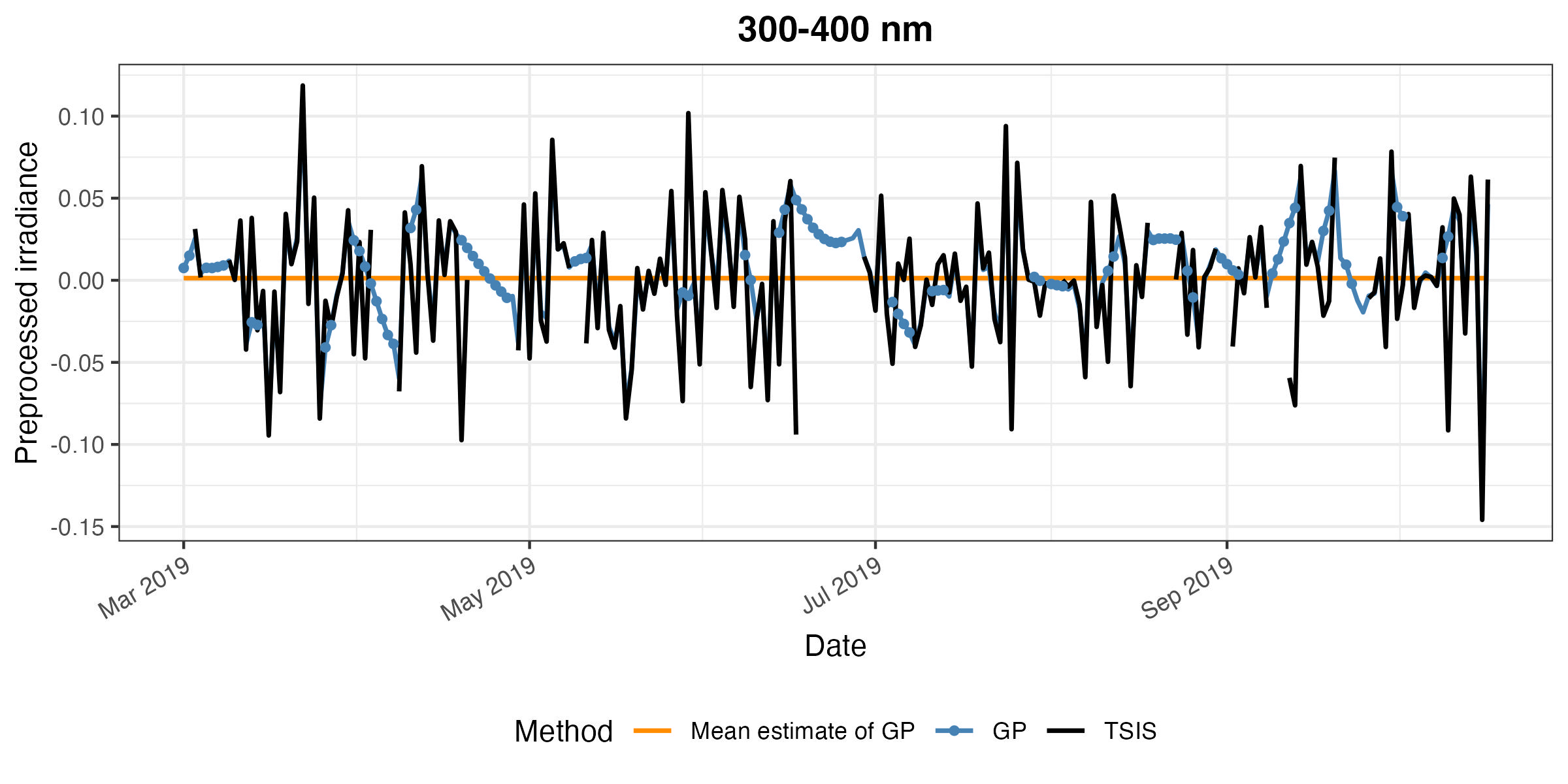}
    \caption{GP imputation on preprocessed TSIS-1 data in the 300–400 nm band. The GP mean estimate (orange) is flat near zero. GP-imputed values and preprocessed TSIS-1 observations are shown in blue and black, respectively. Solid points mark the imputed entries during downtime. }
    \label{fig:gp raw}
\end{figure}

\begin{figure}[t]
    \centering
    \begin{subfigure}[b]{0.9\linewidth}
        \centering
        \includegraphics[width=\linewidth]{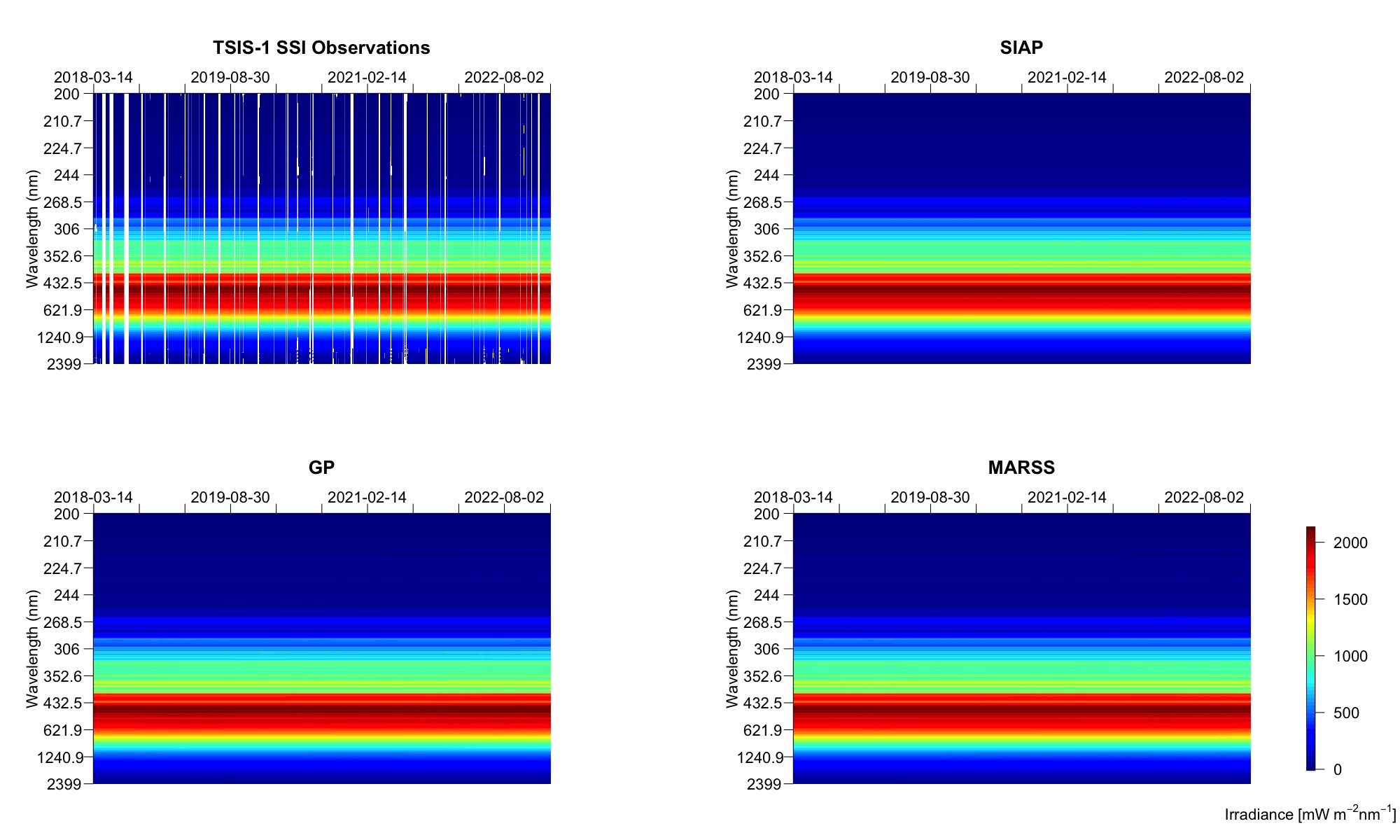}
        \caption{Snapshot of SSI reconstruction.}
        \label{fig:snapshot}
    \end{subfigure}
    \begin{subfigure}[b]{0.9\linewidth}
        \centering
        \includegraphics[width=\linewidth]{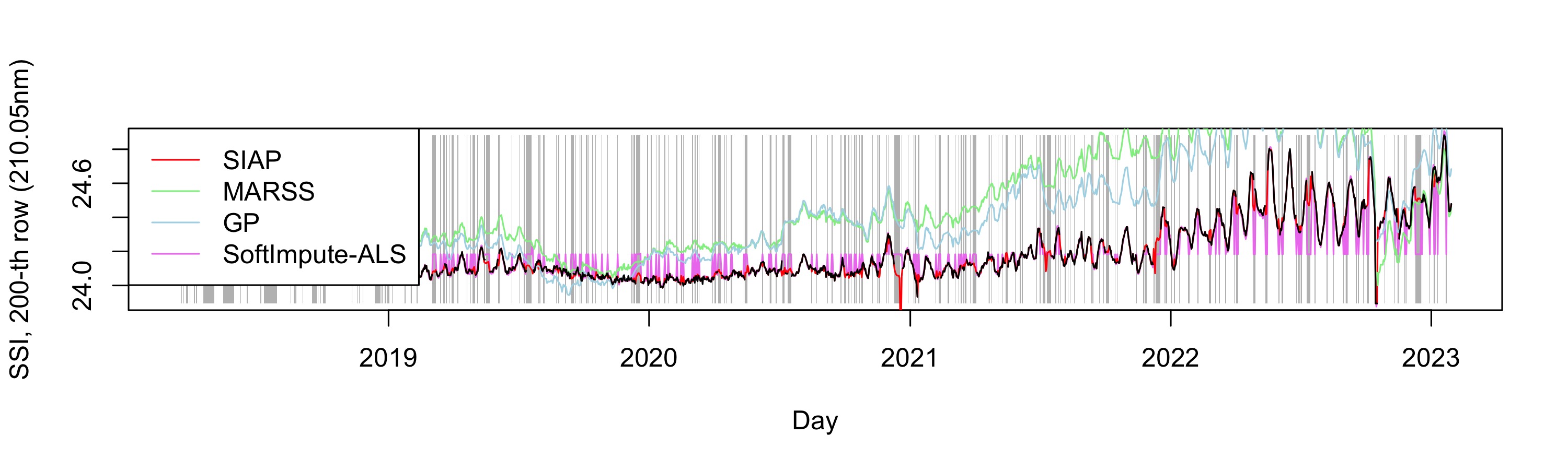}
        \caption{Temporal dimension comparison. Downtime missingness are marked as the gray shaded area.}
        \label{fig:timeseries}
    \end{subfigure}
    \begin{subfigure}[b]{0.95\linewidth}
        \centering
        \includegraphics[width=\linewidth]{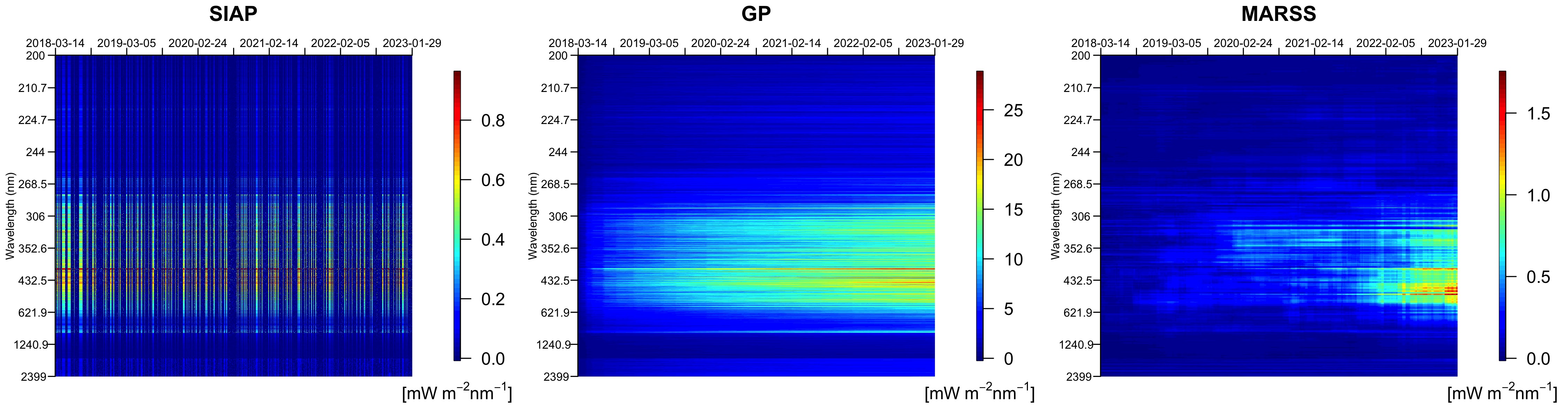}
        \caption{The half width of the pixel-wise uncertainty intervals.}
        \label{fig:uq}
    \end{subfigure}
    \caption{Visualization of SSI reconstruction.}
    \label{fig:snapshot and timeseries}
\end{figure}

Supplementary experiment results are also included in this section. Figure~\ref{fig:alpha path} showcases the change of cross-validation error as $\alpha$ increases.
Figure~\ref{fig:snapshot and timeseries} visualizes the reconstruction results.
Figure~\ref{fig:gp raw} shows raw GP output against preprocessed TSIS-1 data in the 300–400 nm band, runs of positive observed values produce runs of positive imputations. After back-transformation, this behavior produces the upward-trending GP curve.


\bibliographystyle{template_aoas/imsart-nameyear}
\bibliography{ref}